\documentclass[notoc]{JHEP3}
\usepackage{graphicx}
\usepackage{amsmath}
\usepackage{psfrag}
\usepackage{slashed}
\usepackage[sort&compress,numbers]{natbib}

\def\la{\langle}
\def\ra{\rangle}

\def\A#1#2{\la#1#2\ra}
\def\B#1#2{[#1#2]}
\def\AB#1#2#3{\la#1|#2|#3]}
\def\BA#1#2#3{[#1|#2|#3\ra}
\def\AA#1#2#3{\la#1|#2|#3\ra}
\def\BB#1#2#3{[#1|#2|#3]}

\def\e{\epsilon}

\def\fl#1{#1^\flat}
\def\kf#1{K^\flat_{#1}}
\def\bl{\bar{l}}
\def\tl{l_{[-2\e]}}
\def\gg#1#2{\gamma_{#1#2}}
\def\gammab{\bar{\gamma}}
\def\inf{{\rm Inf}}
\def\res{{\rm Res}}

\preprint{IPhT/t08/106\\January 2009}

\title{Direct Extraction Of One Loop Rational Terms}

\author{S. D. Badger \\
Institut de Physique Th\'{e}orique,
CEA-Saclay, F-91191 Gif-sur-Yvette, France \\
E-mail: {\tt simon.badger@cea.fr}
}

\abstract{We present a method for the direct extraction of rational contributions to one-loop scattering
amplitudes, missed by standard four-dimensional unitarity techniques. We use generalised unitarity
in $D=4-2\e$ dimensions to write the loop amplitudes in terms of
products of massive tree amplitudes. We find that the rational terms in $4-2\e$ dimensions can be determined
from quadruple, triple and double cuts without the need for independent pentagon contributions using
a massive integral basis. The additional mass-dependent integral coefficients may then be extracted
from the large mass limit which can be performed analytically or numerically. We check the method by computing the rational parts of all
gluon helicity amplitudes with up to six external legs. We also present a simple application to
amplitudes with external massless fermions.
}


\begin{document}

\section{Introduction}

The main goal of the Large Hadron Collider, due to start experiments later this year, is to explore
the electroweak symmetry-breaking scale and search for new physical phenomena at the TeV scale. In
order to effectively achieve this goal it will be necessary to have precise predictions of
backgrounds within the Standard Model and particularly for the enormous number of QCD or
QCD-associated events that could mask the effects of new particles.

At the present time much progress has been made towards completing calculations for important
cross sections \cite{Bern:NLOwg}. For a selected number of processes, NNLO precision may be
necessary but for the majority of events it is expected that NLO predictions will be sufficient.
Because generic signals of new
phenomena will be associated with the production and subsequent decay of heavy particles, the main
source of backgrounds comes from multi-jet final states. One of the most important ingredients for
such multiparticle NLO cross sections is the virtual matrix elements.  The traditional Feynman
approach to such calculations is extremely computationally intensive due to the rapid growth in the
number of diagrams with the number of external legs.

On-shell techniques, pioneered by Bern, Dixon and Kosower during the mid-nineties
\cite{Bern:uni1,Bern:uni2}, offer an elegant alternative to the traditional Feynman approach. When
constructing loop amplitudes from on-shell objects one works only with the physical degrees of
freedom which can substantially reduce the complexity of the calculation. Loop amplitudes are
re-constructed by sewing together products of on-shell tree amplitudes to extract information about
the branch cuts in each channel of the momentum invariants. 
This information can then be used to find the coefficients of the known scalar integral basis
\cite{Ellis:ints,tHooft:ints,Denner:boxint}. In supersymmetric theories
performing the on-shell cut in four dimensions is sufficient to reconstruct the full amplitude,
whereas in QCD we will be left with additional rational terms. Knowledge of universal factorisation
and use of triple as well as double cuts \cite{Bern:Zto4p} can be used to supplement this approach
and reconstruct full amplitudes in some cases to all multiplicity. 

More recently on-shell techniques have seen renewed interest, via excursions into twistor space,
through developments exploiting the use
of complex momenta. On-shell recursive techniques at tree-level \cite{Britto:rec1,Britto:rec2} use
basic complex analysis and universal factorisation properties to write simple relations between
on-shell amplitudes. These relations have been used successfully to derive compact analytic expressions for a
wide variety of processes. Since the use of complex momenta ensures that three-point amplitudes are
well defined on-shell, it is also possible to refine the generalised unitarity procedure and isolate
individual coefficients of the integral basis \cite{Britto:genu}. Using spinor
integration techniques \cite{Britto:sqcd,Mastrolia:3cut} it has been possible to derive analytic expressions for the cut-constructible
parts of all six-point gluon amplitudes \cite{Britto:sqcd,Britto:N0}. Such techniques are not
restricted to massless theories and have also been developed for massive theories. They can also be
applied to $D$-dimensional cuts
\cite{Brandhuber:gu,Brandhuber:moregu,Anastasiou:dduni1,Anastasiou:dduni2,Britto:intcoeffs,Britto:massuni,Britto:1lmassgu},
with recent applications to some examples of complete gluon amplitudes
\cite{Britto:ddmassless,Feng:spurious}.

Given that the integral coefficients are rational functions of the external momenta it makes sense
to look for a purely algebraic procedure which avoids any explicit integration.
Cutting four propagators in four dimensions completely freezes the loop integral which allows
integration to be replaced by
algebraic operations \cite{Britto:genu}. After integrating a generic triple cut over the on-shell delta functions a one
dimensional integral still remains and it is difficult to separate any new independent functions
from the residues of previously computed, higher order poles. Ossola,
Papadopoulos and Pittau showed that by a special parametrisation of the loop momentum one can find a
systematic way to compute these higher order terms \cite{Ossola:OPP}. After a subtraction of such
terms at the integrand level the remaining terms can be found by solving an algebraic system of
linear equations. This
method has been shown to be successful in the context of Feynman diagram calculations
\cite{Binoth:NLOWWW,Mastrolia:OPPopt,Ossola:rational} and have been implemented in a public code
\texttt{CutTools}
\cite{Ossola:CutTools}. 

A powerful analytical approach is the one by Forde who used simple complex analysis
to go beyond the OPP approach and isolate the coefficients of the scalar integrals \cite{Forde:intcoeffs}. Introducing a complex
parametrisation for the loop momentum, the coefficients are then completely determined
through the limiting behaviour of the products of tree amplitudes. In particular this avoids the
need to solve an algebraic system of equations and leads directly to compact
analytic expressions for the coefficients. One can understand this procedure
further by observing that the non-trivial integrations are simply contour integrals in the complex
plane which, after application of Cauchy's theorem, give compact descriptions of triangle
coefficients \cite{BjerrumBohr:3masstri}. This method has been generalised to accommodate arbitrary
internal masses \cite{Kilgore:massuni} and, with modifications to remove higher order poles from the
complex plane, shown to be an efficient numerical tool \cite{Berger:blackhat}. The technique has
also shown to be a powerful tool for analysing non-trivial cancellations in gravitational theories
\cite{Bern:gravcancel} and for analytic computations of multi-photon amplitudes
\cite{Bernicot:6photon,Bernicot:massqed}.

Two alternative approaches to the computation of the rational terms have been followed in recent
years. The first of these is the use of loop-level recursion relations exploiting the multi-particle factorisation
properties \cite{Bern:1lrec1,Bern:1lrec2,Bern:bootstrap}. This method has been used successfully to derive
analytic expressions for many helicity amplitudes up to eight final state gluons \cite{Berger:genhels},
as well as all-multiplicity expressions for one-loop MHV amplitudes
\cite{Forde:1lmhv,Berger:1lmhv}. The method applies
equally well to amplitudes with massive external particles where amplitudes with gluons coupling to
a Higgs boson have also been derived \cite{Berger:higgsfinite,Badger:1lhiggs,Glover:1lhiggsgen}. A
good review of these new techniques together can be found in reference \cite{Bern:review}. These
recursion relations have been combined with the coefficient extraction of Forde to produce an
automated \texttt{C++} code, {\tt BlackHat}
\cite{Berger:blackhat}. Computation of amplitudes with up to eight external gluons have shown
promising speed and accuracy at fixed precision.

The second method is to use $D$-dimensional cutting techniques which also completely determine the
loop amplitude \cite{Bern:massuni,Bern:selfdualYM}. Recently much progress has been made towards
a numerical approaches to these techniques, based on the OPP approach. This has been shown to be
much faster \cite{Ellis:numuni} than the current Feynman based techniques
\cite{Binoth:golem1,Binoth:golemrat,Giele:num,Ellis:num,Ellis:6ptnum}.
Giele, Kunszt and Melnikov \cite{Giele:dduni} have used an approach using higher integer dimensions
to provide a purely numerical procedure for the evaluation of the $D$-dimensional coefficients. This has been implemented
in a \texttt{Fortran} code, \texttt{Rocket}, and used to compute gluon
amplitudes with up to 20 external legs \cite{Giele:rocket}. Very recently this method has also been applied to
amplitudes with massive fermions \cite{Ellis:massdduni}. Ossola,
Papadopoulos and Pittau have also proposed a technique to calculate these rational parts
\cite{Ossola:rational} using a massive integral basis which motivates the construction we present
here.

The purpose of this paper is to extend the current $D$-dimensional approaches to computations of
rational terms by making use the complex analysis techniques used by
Forde for the cut-constructible terms \cite{Forde:intcoeffs}. This allows us to
compute analytic expressions for full one-loop amplitudes within a single framework and sheds further
light on the efficient computations of full one-loop amplitudes. Just as in Forde's
analysis, this avoids the need to solve an algebraic system of equations present in the OPP approach
and leads directly to compact analytic expressions. The main simplification arising from this
analysis is the separation of pentagon contributions, which vanish in the four-dimensional limit and
so are eliminated from the computation. This leaves the rational terms in
terms of tree amplitudes evaluated in the large-mass limit of box, triangle and bubble cuts. Alternatively this procedure can be
understood as a contour integration for a complex mass parameter where the radius of the contour is
taken to infinity.
For the main part of this paper we concentrate on amplitudes with external gluons though the methods presented
should also apply to more general external
states by using the full $D$-dimensional tree amplitudes. This is demonstrated using a simple
example involving massless external fermions. In section \ref{sec:dints} we review the
$D$-dimensional integral basis and the general form of the rational terms. In section
\ref{sec:extract} we describe how each of the components of the $4-2\e$-dimensional basis can be determined from
the large mass behaviour of four-dimensional, massive, generalised cuts. Section \ref{sec:gluons}
we present some analytic examples of gluon amplitudes with up to six external legs and outline a simple
numerical implementation. We then present a simple application to a four-point massless fermion
amplitude in section \ref{sec:quarks} before giving our conclusions. Some additional notes on mass dependence of the one-loop
integrands and general forms of the boundary expansions are given in an appendix.

\section{Notation}

Throughout this paper we will be considering colour-ordered helicity amplitudes. We use the standard
spinor-helicity
formalism to describe all momenta and external wavefunctions
\cite{Berends:spinhel,deCausmaeker:spinhel,Gunion:spinhel,Xu:spinhel,Kleiss:massspin,Hagiwara:massspin}.
Two component Weyl spinors are written as
\def\tlambda{\tilde{\lambda}}
\begin{align}
	\lambda_\alpha(p) = |p\ra, \qquad \tlambda^{\dot{\alpha}}(p) = |p].
	\label{eq:spinors}
\end{align}
The indices of the two-component spinors are raised and lowered using two totally anti-symmetric
tensors $\e_{\alpha\beta}$ and $\e_{\dot{\alpha}\dot{\beta}}$ where $\e_{12}=\e_{\dot{1}\dot{2}}=1$.
The scattering amplitudes are then written in terms of spinor products defined by,
\begin{align}
	\lambda^\alpha(p)\lambda_\alpha(q) = \A pq, \qquad
	\tlambda_{\dot\alpha}(p)\tlambda^{\dot\alpha}(q) = \B pq.
	\label{eq:spinorprods}
\end{align}
Throughout the paper all momenta are written in matrix form via contraction with the Pauli
$\sigma$ matrices:
\begin{align}
	p^\nu &= \AB{p}{\sigma^\nu}{p}, &
	p\cdot\sigma_{\alpha\dot\alpha} &= |p\ra_{\alpha}[p|_{\dot{\alpha}},
	\label{eq:mommatrix}
\end{align}
where we use a shorthand notation,
\begin{align}
	p &\equiv |p\ra[p|.
\end{align}
For numerical evaluation of the spinor products we have used the standard approach as outlined in
reference \cite{Dixon:TASI}. Modifications to accommodate complex and massive momenta have recently
been implemented in a Mathematica package \cite{Maitre:SAM}.

\section{$D$-dimensional cuts and rational terms \label{sec:dints}}

In this section we review the $D$-dimensional integral basis as considered in previous constructions
\cite{Giele:dduni,Ossola:rational,Anastasiou:dduni1}. Here we focus on the connection between the
$4-2\e$ dimensional representation and that associated with an effective mass $\mu^2$
\cite{Mahlon:1lgluon,Bern:massuni,Bern:selfdualYM}.

We begin by writing a general 1-loop amplitude in terms of a $D$-dimensional $n$-point function,
\begin{align}
	A_n^{(1)}=&\int \frac{d^D l}{(4\pi)^{D/2}} 
	\frac{ {\cal N}(\{p_i\},l)}{(l^2-m_1^2)( (l-K_1)^2-m_2^2)\ldots((l+K_n)^2-m_n^2)}.
	\label{eq:1lgentens}
\end{align}
The numerator function $\cal N$ contains all information from external polarisation states and
wavefunctions and tensor structures from the loop momenta. Since we are concerned with computing the
four dimensional limit it is useful to decompose the loop momenta as,
\begin{equation}
	l^\nu = \bl^\nu + \tl^\nu,
	\label{eq:ddldecomp}
\end{equation}
where $\bl$ contains the four-dimensional components and $\tl$ contains the remaining $D-4=-2\e$ dimensional
components. Using $D$-dimensional Passarino-Veltman reduction techniques on \eqref{eq:1lgentens}
allows us to reduce to a basis of scalar integral functions with rational, but $D$-dimensional, coefficients
\cite{Giele:dduni},
\begin{align}
	&A_n^{(1),D} = 
	\sum_{K_5} \tilde{\mathcal{C}}_{5;K_5}(D)I_{5;K_5}^D
	+\sum_{K_4} \tilde{\mathcal{C}}_{4;K_4}(D)I_{4;K_4}^D\nonumber\\
	+&\sum_{K_3} \mathcal{C}_{3;K_3}(D)I_{3;K_3}^D
	+\sum_{K_2} \mathcal{C}_{2;K_2}(D)I_{2;K_2}^D
	+ \mathcal{C}_{1}(D)I_{1}^D,
	\label{eq:ddintbasis1}
\end{align}
where we define the sets of external momenta, $K_r$, as the set of all ordered partitions of the $n$
external particles into $r$
distinct groups (the ordering is defined by that of the full amplitude $A_n^{(1)}$). We
proceed by writing the amplitude in terms of an integral basis with $D$ independent coefficients
at the cost of expanding the basis of
integral functions. Working in the four dimensional helicity (FDH) scheme\footnote{It is
straightforward to convert gluon amplitudes into the 't Hooft Veltman scheme by subtracting a factor of
$\tfrac{c_\Gamma}{3}A_{\rm tree}$ from the result in the FDH scheme. Similar relations exist for
amplitudes with external fermions \cite{Bern:qqggg}.}
\cite{Bern:1lFDH,Bern:2lFDH} we keep all the external
momenta and sources in four dimensions which means that the dependence on $D$ can only arise through contracting
the loop momentum with itself,
\begin{equation}
	l^2 = \bl^2+\tl^2\equiv \bl^2-\mu^2.
\end{equation}
We can now interpret all dimensional dependence of the coefficients in terms of their dependence on
$\mu^2$. We also know that in any renormalisable gauge theory the maximum rank of an $n$-point tensor integral
appearing in the amplitude is $n$, hence for the box functions we can have up to a maximum
power of $\mu^4$ in the coefficient and up to $\mu^2$ in the triangles and bubbles. The pentagon
integral is only
an independent function in $D$ dimensions since we can find poles in the $D-4$ dimensional
sub-space. As a result, the coefficient of this function in $D=4-2\e$, or residue around the extra dimensional
poles, can have no dependence on $\e$. Therefore we arrive at a new basis,
\begin{align}
	&A_n^{(1),D} = 
	\sum_{K_5} \tilde C_{5;K_5}I_{5;K_5}^D \nonumber\\
	&+\sum_{K_4} C_{4;K_4}^{[0]}I_{4;K_4}^D[1]
	+\sum_{K_4} C_{4;K_4}^{[2]}I_{4;K_4}^{D}[\mu^2]
	+\sum_{K_4} C_{4;K_4}^{[4]}I_{4;K_4}^{D}[\mu^4]
	+\sum_{K_3} C_{3;K_3}I_{3;K_3}^D[1]\nonumber\\
	&+\sum_{K_3} C_{3;K_3}^{[2]}I_{3;K_3}^{D}[\mu^2]
	+\sum_{K_2} C_{2;K_2}I_{2;K_2}^D[1]
	+\sum_{K_2} C_{2;K_2}^{[2]}I_{2;K_2}^{D}[\mu^2]
	+ C_{1}I_{1}^D.
	\label{eq:ddintbasis2}
\end{align}
The integrals over $\mu^2$ can be performed by separating the integration into $4$ and $D-4$
dimensional parts,
\begin{equation}
	\int \frac{d^Dl_1}{(2\pi)^D} = \int \frac{d^{-\e}(\mu^2) }{(2\pi)^{-2\e}}\int
	\frac{d^4\bl_1}{(2\pi)^4}.
\end{equation}
It is fairly straightforward to write the four new integrals in terms of higher-dimensional scalar
integrals using \cite{Bern:selfdualYM,Bern:massuni},
\begin{equation}
		I_n^D[\mu^{2r}] = \frac{1}{2^r} I_n^{D+2r}[1] \prod_{k=0}^{r-1} (D-4+k). 
\end{equation}
We also use the dimensional shift identity \cite{Bern:pentagon} to decompose the pentagon
integrals\footnote{We refer the reader the \cite{Bern:pentagon} for full definitions of the
quantities in eq. \eqref{eq:pentred} although we point out that all integrals include a factor of
$(-1)^{n+1}$ in their definition. The essential information is that the coefficients are just
functions of the external momenta and the internal masses.}
\begin{align}
	\label{eq:pentred}
	I^D_5[1] &= \frac{(D-4)}{2} I^{D+2}_5[1] \left(\sum_{i,j} S^{-1}_{ij}\right) +
	\frac{1}{2}\sum_{i=1}^5\sum_{j} S^{-1}_{ij} I_{4;K_5^{(i)}},\\
	S_{ij} &= \frac{1}{2}\left( m_i^2+m_j^2-p_{ij}^2 \right).
\end{align}
In the above $K_5^{(i)}$ is one of the five sets of four partitions obtained cyclically merging two
adjacent partitions of a given pentagon configuration $K_5$. 
After this is done the explicit $D$-dependence of the amplitude is restored \cite{Giele:dduni}:
\begin{align}
	&A_n^{(1),D} = 
	\frac{D-4}{2}\sum_{K_5} C_{5;K_5}I_{5;K_5}^{D+2} \nonumber\\
	&+\sum_{K_4} C_{4;K_4}I_{4;K_4}^D
	+\frac{D-4}{2}\sum_{K_4} C_{4;K_4}^{[2]}I_{4;K_4}^{D+2}
	+\frac{(D-4)(D-2)}{4}\sum_{K_4} C_{4;K_4}^{[4]}I_{4;K_4}^{D+4}
	+\sum_{K_3} C_{3;K_3}I_{3;K_3}^D\nonumber\\
	&+\frac{D-4}{2}\sum_{K_3} C_{3;K_3}^{[2]}I_{3;K_3}^{D+2}
	+\sum_{K_2} C_{2;K_2}I_{2;K_2}^D
	+\frac{D-4}{2}\sum_{K_2} C_{2;K_2}^{[2]}I_{2;K_2}^{D+2}
	+ C_{1}I_{1}^D,
	\label{eq:ddintbasis3}
\end{align}
where
\begin{align}
	C_{4;K_4} &= C_{4;K_4}^{[0]} + \sum_{i=1}^5\sum_{j} S^{-1}_{ij} \tilde C_{5;K_5^{(i)}}.
	\label{eq:4dpentred}\\
	C_{5;K_5} &= \tilde{C}_{5;K_5}\sum_{i,j} S^{-1}_{ij}.
	\label{eq:ddpentcf}
\end{align}
After taking the 4-dimensional limit, $D\to4-2\e$, we find that the integral basis reduces to a
combination of box, triangle and bubble integrals but at the cost of introducing additional rational
terms,
\begin{equation}
	A_n^{(1),4-2\e} = 
	\sum_{K_4} C_{4;K_4}I_{4;K_4}^{4-2\e}
	+\sum_{K_3} C_{3;K_3}I_{3;K_3}^{4-2\e}
	+\sum_{K_2} C_{2;K_2}I_{2;K_2}^{4-2\e}
	+C_{1}I_{1}^{4-2\e}+R_n+\mathcal{O}(\e).
	\label{eq:4-2eintbasis}
\end{equation}
The final step is to identify how the higher dimensional integrals in \eqref{eq:ddintbasis3}
contribute to the rational terms. This is surprisingly simple since the scalar box and scalar
pentagon integrals are finite in $6-2\e$ dimensions and don't contribute to the rational part. The
remaining three terms, written in terms of
integrals over $\mu$, are,
\begin{align}
	I_4^{4-2\e}[\mu^4] &\overset{\e\to0}{\to} -\frac{1}{6}, \nonumber\\
	I_3^{4-2\e}[\mu^2] &\overset{\e\to0}{\to} -\frac{1}{2}, \nonumber\\
	I_2^{4-2\e}[\mu^2] &\overset{\e\to0}{\to} -\frac{1}{6}\left( s - 3(m_1^2+m_2^2) \right).
	\label{eq:muints}	
\end{align}
The rational terms are thus given by \cite{Giele:dduni,Ossola:rational,Britto:1lmassgu},
\begin{equation}
	R_n = -\frac{1}{6} \sum_{K_4} C_{4;K_4}^{[4]}
	-\frac{1}{2} \sum_{K_3} C_{3;K_3}^{[2]}
	-\frac{1}{6} \sum_{K_2} \left( K_2^2-3(m_1^2+m_2^2) \right)C_{2;K_2}^{[2]}	
	\label{eq:rational}
\end{equation}

\section{Extracting the integral coefficients using massive propagators \label{sec:extract}}

To extract the integral coefficients using generalised unitarity we need to solve the constraints
which put the various propagators on-shell \cite{Forde:intcoeffs}. To generalise this to the
$D$-dimensional case we also need to extract the $\mu$ dependence of the coefficients as defined in eq.
\eqref{eq:ddintbasis2}. Since the extra dimensions in the loop momentum can be interpreted as an
effective mass term, it is possible to construct the full amplitude from tree amplitudes where the
internal legs have a uniform mass:
\begin{equation}
	l_i^2 = \bl_i^2-\mu^2 = 0 \Rightarrow \bl_i^2=\mu^2.
\end{equation}
This method has been used successfully within the standard unitarity cut technique
\cite{Bern:massuni,Bern:selfdualYM} and in conjunction
with spinor integration \cite{Britto:1lmassgu,Britto:ddmassless,Feng:spurious}.
Solving the system of on-shell constraints can then be achieved in exactly the same way as the four-dimensional massive case
\cite{Kilgore:massuni}. For the current study we also consider ourselves to be restricted to cases
with $D$-dimensional scalars with massless external fermions and gauge bosons.

\subsection{Box Coefficients}

\begin{figure}[h]
	\begin{center}
		\psfrag{K1}{$K_4$}
		\psfrag{K2}{$K_1$}
		\psfrag{K3}{$K_2$}
		\psfrag{K4}{$K_3$}
		\psfrag{l}{$l_1$}
		\psfrag{l-K2}{\hspace{7mm}$l_2$}
		\psfrag{l-K2-K3}{$l_3$}
		\psfrag{l+K1}{\hspace{7mm}$l_4$}
		\includegraphics[width=5cm]{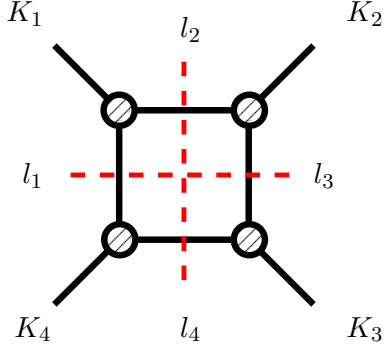}
	\end{center}
	\caption{A general quadruple cut with loop momentum flowing clockwise and all external
	momenta outgoing.}
	\label{fig:box}
\end{figure}

In this section we will show that by extracting the coefficient of the $D$-dimensional box directly using Forde's formalism
\cite{Forde:intcoeffs}, we can ignore the pentagon coefficients entirely. We begin by
choosing the four-momentum, $\bl_1$, to be parametrised by,
\begin{align}
	\bl_1 &= a\kf4+b\kf1+c|\kf4\ra[\kf1|+d|\kf1\ra[\kf4|.
\end{align}
where the $\kf {1,4}$ define a massless basis in terms of two of the external momenta:
\begin{align}
	\kf 4 &= \frac{\gg 14\left(\gg14 K_4-S_4 K_1\right)}{\gg 14^2-S_1S_4}, &
	\kf 1 &= \frac{\gg 14\left(\gg14 K_1-S_1 K_4\right)}{\gg 14^2-S_1S_4}, &
	\gg 14 &= K_1\cdot K_4 \pm \sqrt{(K_1\cdot K_4)^2-S_1S_4}, &
	\label{eq:flatdef}
\end{align}
for $S_i = K_i^2$.

The four on-shell constraints then fix the coefficients:
\begin{align}
	\bl_1 &= a\kf4+b\kf1+c|\kf4\ra[\kf1|+\frac{\gg 14 ab-\mu^2}{c\gg 14}|\kf1\ra[\kf4|\nonumber\\
	    &= \bl_1^\flat - \frac{\mu^2}{c\gg 14}|\kf1\ra[\kf4|
\end{align}
where
\begin{align}
	a=\frac{S_1(S_4+\gg 14)}{\gg 14^2-S_1S_4}, \qquad&\qquad
	b=-\frac{S_4(S_1+\gg 14)}{\gg 14^2-S_1S_4}, \qquad\qquad
	c_\pm = \frac{-c_1\pm\sqrt{c_1^2-4c_0c_2}}{2c_2}, \label{eq:csol}\\
	c_2 &= \AB{\kf 4}{K_2}{\kf 1}, & & \\
	c_1 &= a \AB{\kf 4}{K_2}{\kf 4}+b \AB{\kf 1}{K_2}{\kf 1}-S_2-2K_1\cdot K_2, \\
	c_0 &= \left(ab-\frac{\mu^2}{\gg 14}\right)\AB{\kf 1}{K_2}{\kf 4}.
\end{align}
For the quadruple cut we find that both solutions for $\gg 14$ are degenerate.

Now consider the quadruple cut:
\begin{align}
	&(4\pi)^{D/2}\int \frac{d^Dl_1}{(2\pi)^D} (-2\pi i)^4 \prod_{i=1}^{4} \delta(l_i^2) A_1 A_2 A_3 A_4 \nonumber\\
	= &(4\pi)^{D/2}\int \frac{d^{-2\e}\mu }{(2\pi)^{-2\e}} \int d^4\bl_1 \prod_{i=1}^{4}
	\delta(\bl_i^2-\mu^2)
	A_1 A_2 A_3 A_4 \nonumber\\
	=& (4\pi)^{D/2}\int \frac{d^{-2\e}\mu }{(2\pi)^{-2\e}} \sum_{\sigma}\left[
	\inf_{\mu^2}[A_1A_2A_3A_4(\bl_1^\sigma)]
	+ \sum_{ {\rm poles} \{i\}} \frac{ \res_{\mu^2=\mu^2_i} (A_1 A_2 A_3
	A_4(\bl_1^\sigma))}{\mu^2-\mu^2_i}\right].
\end{align}
The first term encodes all the information from the boundary of the $\mu$ contour integral. The
$\inf$ operation therefore takes the form of a polynomial in $\mu$ which is cut off at some maximum
power,
\begin{equation}
	\inf_{\mu^2}[f(\mu)] = \sum_{k=0}^{p} c_k \mu^{2k}
\end{equation}
The second term contains information about the pentagon coefficients since it has
an extra propagator in $\mu^2$ we can identify these terms with coefficients in the basis by
performing a partial fractioning in $\mu^2$ and comparing with equation \eqref{eq:ddintbasis3}:
\begin{align}
	&\frac{{\rm Res}_{\mu^2=\mu^2_i}(A_1A_2A_3A_4(\bl_1^\sigma))}{\mu^2-\mu_i^2} 
	= \nonumber\\ 
	&\hspace{1cm}\frac{\mu^2{\rm Res}_{\mu^2=\mu^2_i}(A_1A_2A_3A_4(\bl_1^\sigma))}{\mu_i^2(\mu^2-\mu_i^2)} 
	-\frac{{\rm Res}_{\mu^2=\mu^2_i}(A_1A_2A_3A_4(\bl_1^\sigma))}{\mu_i^2}.
\end{align}
The first term in the last equation gives the contribution to the $D+2$ dimensional pentagon
integral whereas the second term combines with the $\mu^0$ component of the boundary term to form
the coefficient of the $D=4-2\e$ dimensional box integral. It is true in any case that neither of
these terms contribute to the rational part and therefore we can extract the information from
standard four dimensional cuts. Explicitly we find,
\begin{equation}
	\int d^{-2\e}\mu \frac{\mu^2\res_{\mu^2=\mu_i^2}(A_1A_2A_3A_4)}{\mu_i^2(\mu^2-\mu_i^2)}
	\overset{\e\to0}{\to} 0,
\end{equation}
which matches up with the vanishing of the $D+2$ dimensional pentagons from equation
\eqref{eq:ddintbasis3}.

The rational contribution is therefore found by looking at the behaviour of the product of the four
tree amplitudes at large values of $\mu$,
\begin{align}
	\label{eq:ddbox1}
	\frac{i}{2}\sum_{\sigma=\pm}\inf_{\mu^2}[A_1 A_2 A_3 A_4(\bl_1^\sigma)] =
	\sum_{k=0}^{4} \mu^k C_4^{[k]} \\
	\Rightarrow C_4^{[4]} = \frac{i}{2}\sum_{\sigma=\pm}\inf_{\mu^2}[A_1 A_2 A_3
	A_4(\bl_1^\sigma)]|_{\mu^4},
	\label{eq:ddbox}
\end{align}
where we have remembered that the quadruple cut of the scalar box integral is $-i(4\pi)^{D/2}$. In
the final formula, the
$\inf_{\mu^2}$ operation has been restricted to the coefficient of the $\mu^4$ term of the
polynomial. We discuss the maximum possible power of $\mu^2$ appearing in eq. \eqref{eq:ddbox1} in Appendix
\ref{app:mucount}. It is straightforward to look at the limit
$\mu\to \infty$ by performing a Taylor expansion to give analytic expressions. 

\subsection{Triangle Coefficients}

\begin{figure}[h]
	\begin{center}
		\psfrag{l1}{$l_1$}
		\psfrag{l2}{$l_2$}
		\psfrag{l3}{$l_3$}
		\psfrag{K1}{$K_3$}
		\psfrag{K2}{$K_1$}
		\psfrag{K3}{$K_2$}
		\includegraphics[width=4cm]{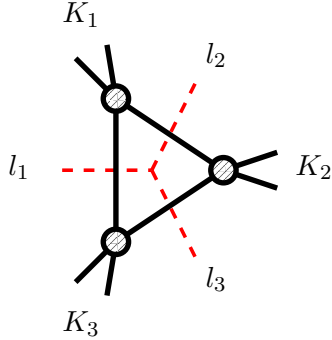}
	\end{center}
	\caption{Momentum conventions for the triple cut; all momenta are outgoing and the loop
	momentum flows clockwise.}
	\label{fig:tri}
\end{figure}

We next consider the triple cut integrals\footnote{We have suppressed the factors of $i$ and $\pi^2$
in the triple and double cuts, these are implicitly put back in when writing the formulae for the
coefficients.} which, with the new dependence on $\mu$, can be written as,
\begin{align}
	& \int d^{-2\e}\mu \int d^4\bl_1 \prod_{i=1}^{3}
	\delta(\bl_i^2-\mu^2) A_1A_2A_3 \nonumber\\
	=& \int d^{-2\e}\mu \int dt J_t \bigg(
	\inf_{\mu^2}[\inf_t[A_1A_2A_3]] 
	+ \sum_i\inf_{\mu^2}\left[\frac{\res_{t=t_i}(A_1A_2A_3)}{t-t_i}\right]\nonumber\\
	&+ \sum_i\frac{\res_{\mu^2=\mu^2_i}(\inf_t[A_1A_2A_3])}{\mu^2-\mu^2_i}
	+ \sum_{i,j}\frac{\res_{\mu^2=\mu^2_j}(\res_{t=t_i}(A_1A_2A_3))}{(t-t_i)(\mu^2-\mu^2_j)}
	\bigg)
	\label{eq:triplecut}
\end{align}
Defining $\kf{1,3}$ analogously to equation \eqref{eq:flatdef}, we choose a loop momentum basis,
\begin{equation}
	l_1 = a\kf3+b\kf1+t|\kf3\ra[\kf1|+\frac{ab \gg 13-\mu^2}{\gg 13 t}|\kf1\ra[\kf3|,
	\label{eq:tricutbasis}
\end{equation}
which ensures that there all integrals over $t^n$ and $1/t^n$ vanish \cite{Forde:intcoeffs},
\begin{equation}
	\int dt J_t t^n = 0, \qquad \int dt J_t \frac{1}{t^n} = 0.
\end{equation}
The second and fourth terms can be directly associated with the previously calculated scalar box
coefficients. This is due the presence of an additional propagator term and the lack of $t$-dependence
in the numerator. We can also show that the third term does not contribute to rational part by
applying partial fractioning in an analogous way to the box case. Of the two terms obtained, one will be associated with the $6-2\e$ dimensional boxes and the other with the $4-2\e$
triangle coefficient which is cut-constructible.

Therefore the leading $\mu^2$ dependence of the triangle functions can be completely determined through the boundary behaviour:
\begin{align}
	\label{eq:ddtri1}
	C_3^{[2]} &=
	\frac{1}{2}\sum_{\sigma}\inf_{\mu^2}[\inf_t[A_1A_2A_3(\bl_1^\sigma)]|_{t^0}]|_{\mu^2}
\end{align}
One must also sum over the two
solutions, $\sigma$, for the loop momentum. For the massless case when $S_1\neq0$ and $S_3\neq0$ it
is sufficient to sum over the two solutions for $\gg 13$. However, in general it is necessary to sum
over the solution $\bl_1$ given above in eq. \eqref{eq:tricutbasis} and conjugate solutions given
by,
\begin{equation}
	l_1^* = a\kf3+b\kf1+t|\kf1\ra[\kf3|+\frac{ab \gg 13-\mu^2}{\gg 13 t}|\kf3\ra[\kf1|,
	\label{eq:tricutbasis*}
\end{equation}
evaluated at a fixed value of $\gg 13$. This applies equally well to the one and two mass triangles
when one or both of $S_1,S_3$ vanishes and there is only a single solution for $\gg 13$. As for the
box case, we postpone justification of the form of eq. \eqref{eq:ddtri1} to Appendix \ref{app:mucount}.

\subsection{Bubble Coefficients}

\begin{figure}[h]
	\begin{center}
		\psfrag{l1}{$l_1$}
		\psfrag{l2}{$l_2$}
		\psfrag{l3}{$l_3$}
		\psfrag{K1}{$K_1$}
		\psfrag{K2}{$K_3$}
		\psfrag{K3}{$K_2$}
		\includegraphics[width=8cm]{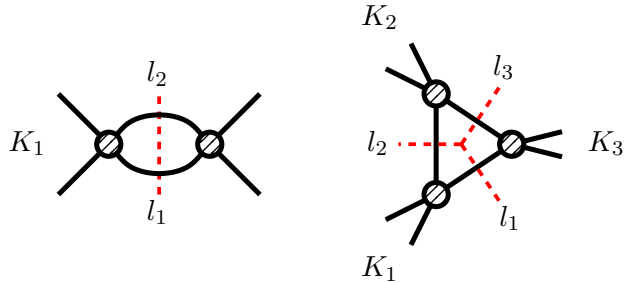}
	\end{center}
	\caption{Pure Bubble and Triangle terms contributing to the bubble coefficients.}
	\label{fig:bub}
\end{figure}

Finally, we find the double cut integral using the following basis for the loop momentum:
\begin{equation}
	\bl_1 = y \kf1 + \frac{S_1(1-y)}{\gammab} \chi + t|\kf1\ra[\chi| + \left( y(1-y)S_1 - \mu^2
	\right) \frac{|\chi\ra[\kf1|}{\gammab t},
\end{equation}
where
\begin{align}
	\kf 1 = K_1 - \frac{S_1	}{\gammab}\chi, \qquad\qquad \gammab=2(K_1\cdot \chi).
	\label{eq:K1def_bub}
\end{align}
The cut integral can then be decomposed into,
\begin{align}
	&\int d^{-2\e}\mu \int d^4\bl_1 \prod_{i=1}^{2}
	\delta(\bl_i^2-\mu^2) A_1A_2 \nonumber\\
	&=\int d^{-2\e}\mu \int dtdy~J_{t,y}~
	\inf_{\mu^2}[\inf_t[\inf_y[A_1A_2]]]\nonumber\\&
	+\sum_{i}\inf_{\mu^2}\left[\inf_t\left[\frac{\res_{y=y_i}(A_1A_2)}{y-y_i}\right]\right]
	+\sum_{i}\inf_{\mu^2}\left[\frac{\res_{t=t_i}(\inf_y[A_1A_2])}{t-t_i}\right]\nonumber\\&
	+\sum_{i,j}\inf_{\mu^2}\left[\frac{\res_{t=t_j}(\res_{y=y_i}(A_1A_2))}{(y-y_i)(t-t_j)}\right]
	+\sum_{i}\frac{\res_{\mu^2=\mu^2_i}(\inf_t[\inf_y[A_1A_2]])}{\mu^2-\mu_i^2}\nonumber\\&
	+\sum_{i,j}\frac{\res_{\mu^2=\mu^2_j}(\inf_t[\res_{y=y_i}(A_1A_2)])}{(\mu^2-\mu_j^2)(y-y_i)}
	+\sum_{i,j}\frac{\res_{\mu^2=\mu^2_j}(\res_{t=t_i}(\inf_y[A_1A_2]))}{(\mu^2-\mu_j^2)(t-t_i)}\nonumber\\&
	+\sum_{i,j,k}\frac{\res_{\mu^2=\mu^2_k}(\res_{t=t_j}(\res_{y=y_i}[A_1A_2]))}{(\mu^2-\mu_k^2)(t-t_j)(y-y_i)}.
	\label{eq:doublecut}
\end{align}
The last four terms contain residues in $\mu$ and as such cannot contribute to the rational terms.
Of these terms, the pure $\mu^2-\mu_i^2$ pole will also have a contribution to the $\mu=0$ (or
cut-constructible) bubble coefficient. The fourth term, in which the numerator is
completely independent of $y$ and $t$, can also be discarded as it has two propagator terms and can
therefore only come from scalar box integrals. The remaining terms do have contributions
to the bubble coefficient, and as described by Forde \cite{Forde:intcoeffs}, can be determined by
computing the $\inf$ expansions for both $y$ and $t$ for both the double cut and triple cuts with
non-vanishing integrals over $t$. The additional $\mu$ dependence is again determined by looking at
the large $\mu$ behaviour:
\begin{align}
	C_2^{ {\rm bub} [2]} &= -i\inf_{\mu^2}\inf_t[\inf_y[A_1A_2(\bl_1(y,t,\mu^2)]]]|_{\mu^2,t^0,y^i\to
	Y_i}\\
	C_2^{ {\rm tri}(K_3) [2]} &=
	-\frac{1}{2}\sum_{\sigma=\pm} 
	\inf_{\mu^2}\inf_t[A_1A_2A_3^{K_3}(\bl_1(y_\sigma,t,\mu^2)]]|_{\mu^2,t^i\to T_i}.
\end{align}
with the full bubble coefficient being a sum of the pure bubble and triangle subtraction terms:
\begin{align}
	C_2^{[2]} = C_2^{ {\rm bub},[2]} + \sum_{\{K_3\}}C_2^{ {\rm tri}(K_3)[2]}.
	\label{eq:ddbub}
\end{align}

The functions $T_i$ and $Y_i$ have been computed recently in references \cite{Forde:intcoeffs,Kilgore:massuni}
for arbitrary kinematics. Explicitly with a uniform internal mass we have,
\begin{align}
	Y_0 = 1 \quad
	Y_1 = \frac{1}{2} \quad
	Y_2 = \frac{1}{3}\left( 1 - \frac{\mu^2}{S_1} \right).
	\label{eq:Ys}
\end{align}
Solving the additional on-shell constraint, $(l_1+K_3)^2=\mu^2$, for the triangle subtraction terms
gives the two solutions for $y$ as,
\begin{align}
	y_\pm = \frac{C_1\pm \sqrt{C_1^2+4C_0C_2}}{2C_2},
	\label{eq:ysols}
\end{align}
where
\begin{align}
	C_2 &= S_1\AB{\chi}{K_3}{\kf 1},\\
	C_1 &= \gammab t \AB{\kf 1}{K_3}{\kf 1}-S_1 t\AB{\chi}{K_3}{\chi} +
	S_1\AB{\chi}{K_3}{\kf 1},\\
	C_0 &= \gammab t^2\AB{\kf 1}{K_3}{\chi} - \mu^2\AB{\chi}{K_3}{\kf 1}
	+ \gammab tS_3 + tS_1\AB{\chi}{K_3}{\chi}.
\end{align}
The non-vanishing integrals over $t$ are given by:
\begin{align}
	\label{eq:Ts1}
	T_1 &= -\frac{S_1\AB{\chi}{K_3}{\kf 1}}{2\gammab\Delta},\\
	\label{eq:Ts2}
	T_2 &= -\frac{3S_1\AB{\chi}{K_3}{\kf 1}^2}{8\gammab^2\Delta^2}\left( S_1 S_3+K_1\cdot
	K_3 S_1\right),\\
	T_3 &= -\frac{\AB{\chi}{K_3}{\kf 1}^3}{48\gammab^3\Delta^3}\bigg( 
	15S_1 S_3^2 + 30K_1\cdot K_3 S_1^3 S_3+
	11 (K_1\cdot K_3)^2 S_1^3 \nonumber\\
	&\hspace{3cm}+ 3S_1^4 S_3 + 16\mu^2 S_1^2 \Delta
	\bigg),
	\label{eq:Ts3}
\end{align}
where $\Delta = (K_1\cdot K_3)^2 - S_1S_3$. 

\section{Rational contributions to gluon amplitudes \label{sec:gluons}}

Integral coefficients in which a single particle type circulates in the loop are particularly
well suited to the method described in the previous section. Such calculations apply to give
the rational terms of the all-gluon amplitudes. The supersymmetric decomposition of such amplitudes is given by,
\begin{align}
	A^g_n &= A_n^{ {\cal N}=4}-4A_n^{ {\cal N}=1}+A_n^{[s]} +N_f\left( A_n^{ {\cal
	N}=1}-A_n^{[s]} \right).
	\label{eq:1lgsuper}
\end{align}
Since supersymmetric amplitudes are cut constructible in four dimensions the rational terms in such an
amplitude only appear in $A_n^{[s]}$:
\begin{equation}
	A_n^{[s]} = C_n^{[s]}+R_n^g.
\end{equation}
We can then compute the rational parts of any scalar
amplitude by introducing an effective mass and evaluating the integral coefficients in four-dimensions
with the tree amplitudes for massive scalars \cite{Badger:mrec,Forde:mscalar,Brandhuber:gu}.

\begin{figure}[t]
	\begin{center}
		\psfrag{i}{\tiny$i$}
		\psfrag{j}{\tiny$j$}
		\psfrag{k}{\tiny$k$}
		\psfrag{l}{\tiny$l$}
		\psfrag{i+1}{\tiny$i+1$}
		\psfrag{j+1}{\tiny$j+1$}
		\psfrag{k+1}{\tiny$k+1$}
		\psfrag{l+1}{\tiny$l+1$}
		\psfrag{C4}{$C_{4;K_{i+1,j}|K_{j+1,k}|K_{k+1,l}}$}
		\psfrag{C3}{$C_{3;K_{i+1,j}|K_{j+1,k}}$}
		\psfrag{C2}{$C_{2;K_{i+1,j}}$}
		\includegraphics[width=12cm]{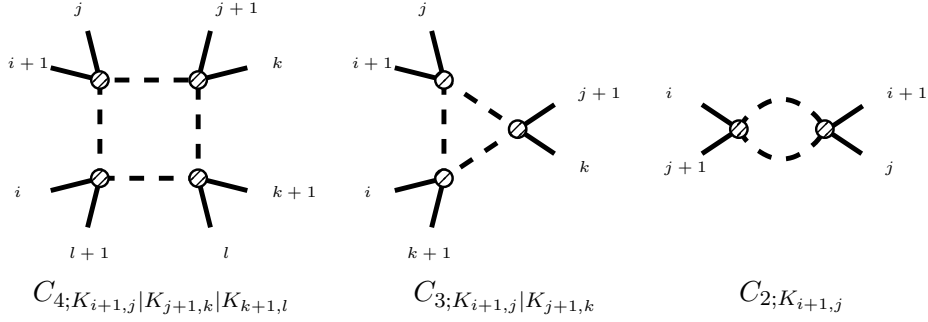}
	\end{center}
	\caption{Cut diagrams for the integral coefficients contributing to the rational parts of
	the $n$-gluon amplitudes.}
	\label{fig:intbasis}
\end{figure}

The rational contribution, written in terms of the massive scalar basis, is therefore,
\begin{align}
	R_n^g = 
	&-\frac{1}{24}\sum_{i=1}^{n}\sum_{j=i+1}^{i-3}\sum_{k=j+1}^{i-2}\sum_{l=k+1}^{i-1}
	C_{4;K_{i+1,j}|K_{j+1,k}|K_{k+1,l}}^{[4]}\nonumber\\
	&-\frac{1}{6}\sum_{i=1}^{n}\sum_{j=i+1}^{i-2}\sum_{k=j+1}^{i-1}C_{3;K_{i+1,j}|K_{j+1,k}}^{[2]}\nonumber\\
	&-\frac{1}{12}\sum_{i=1}^{n}\sum_{j=i+2}^{i-2} s_{i+1,j}\, C_{2;K_{i+1,j}}^{[2]},
	\label{eq:intbasis}
\end{align}
as depicted in figure \ref{fig:intbasis}. The arguments of the integral coefficients are considered
to be taken ${\rm mod}(n)$, which accounts for the extra factors in front of the coefficients due to an
over-counting in the summations. Our notation also suppresses the momentum flowing from one of the
vertices of the integral since it can be inferred from momentum conservation.

\subsection{Analytic expressions for the four-point amplitude}

As an explicit example we present an analytic computation of the four gluon amplitude which has
been considered previously with similar unitarity constructions \cite{Bern:massuni,Britto:1lmassgu}. Here we
only require the three- and four-point tree level amplitudes with a pair of massive scalars
\cite{Badger:mrec,Forde:mscalar}:
\begin{align}
	A_3(1_{s},2^-,3_s) &= i\frac{\AB 21{\xi_2}}{\B2{\xi_2}}, &
	A_3(1_{s},2^+,3_s) &= i\frac{\BA 21{\xi_2}}{\A2{\xi_2}}, \\
	A_4(1_{s},2^+,3^+,4_s) &= i\frac{\mu^2\B 23}{\A23\AB 212}, &
	A_4(1_{s},2^+,3^-,4_s) &= i\frac{\AB 312^2}{s_{23}\AB 212},
	\label{eq:trees}
\end{align}
where $\mu$ is the mass of the scalar particles.
We then construct the integrand for the quadruple cut using $K_4=p_4,K_1=p_1,K_2=p_2$ which leads to a
simplified solution of the general on-shell constraints from eq. \eqref{eq:csol}:
\begin{align}
	a&=0, & b&=0, & d&=-\frac{\mu^2}{c s_{41}}.
\end{align}
The fact that we only require information about the leading behaviour in $\mu^2$ means it is
sufficient to look at the leading behaviour of the coefficient $c$:
\begin{equation}
	c \overset{\mu^2\to\infty}{\to} \pm \mu \sqrt{\frac{\AB{\kf 1}{K_2}{\kf 4}}{\gamma_{14}\AB{\kf
	4}{K_2}{\kf 1}}} = \pm\mu\sqrt{\frac{\AB 124}{\AB421s_{41}}}.
	\label{eq:clim}
\end{equation}
In fact, since we have a product of four tree amplitudes, we find the coefficient is only a
function of $c^2$ and so there are no square roots appearing in the final values:
\begin{align}
	C_4^{[4]}(1^+,2^+,3^+,4^+) &=
\frac{2i[21][43]
}{\la12\ra\la34\ra},\\
	C_4^{[4]}(1^-,2^+,3^+,4^+) &=
-\frac{2i\la14\ra^2[42][43]
}{s_{23}\la24\ra\la34\ra}, \\
	C_4^{[4]}(1^-,2^-,3^+,4^+) &=
\frac{2i\la12\ra[43]
}{\la34\ra[21]}, \\
	C_4^{[4]}(1^-,2^+,3^-,4^+) &=
\frac{2i\la12\ra\la34\ra[42]^2
}{\la24\ra^2[21][43]}.
\end{align}
The triangle coefficients are particularly straightforward to evaluate since the $\inf_t$ operation
is equivalent to performing a Taylor expansion around $t=\infty$. It is helpful to choose the two
massless legs of each triple cut to form the basis since this simplifies the form of the analytic
expression. Let us be slightly more explicit by giving the specific details for the computation of
the $C_{3;12}(1^-,2^+,3^+,4^+)$ coefficient. In this channel it is easiest to choose
$K_1=p_3,K_2=p_4$ since both $S_1$ and $S_2$ will vanish and the two solutions, one each from eq.
\eqref{eq:tricutbasis} and \eqref{eq:tricutbasis*}, are:
\begin{align}
	l_1 &= t|3\ra[4|-\frac{\mu^2}{s_{12}t}|4\ra[3|, \\
	l_1^* &= t|4\ra[3|-\frac{\mu^2}{s_{12}t}|3\ra[4|.
\end{align}
The product of tree amplitudes is trivial to write down:
\begin{align}
	A_1(-l_1,4^+,l_2)A_2(-l_2,1^-,2^+,l_3)A_3(-l_3,3^+,l_1) =
	2i\frac{\AB{1}{l_2}{2}^2\AB{3}{l_1}{4}\AB{4}{l_1}{3}}
	{\A 12 \A 34^2 \B 12 \AB{1}{l_2}{2}}.
	\label{eq:c312integrand}
\end{align}
We are then left to insert the two solutions for $l_1$ and sum over the two values according to
equation \eqref{eq:ddtri}. 
\begin{align}
	\inf_\mu[\inf_t[A_1A_2A_3(l_1)]]|_{\mu^2,t^0} &= -2i\frac{\B 24^2\B 34 s_{23}}{\A 12\A 34\B 12\B 14^2},\\
	\inf_\mu[\inf_t[A_1A_2A_3(l_1^*)]]|_{\mu^2,t^0} &= -2i\frac{\B 24^2\B 34 s_{23}}{\A 12\A 34\B 12\B 14^2}
	\left( 1-\frac{s_{14}^2}{s_{24}^2}\right).
\end{align}
So, after applying momentum conservation, we find the final result:
\begin{equation}
	C_{3;12}^{[2]}(1^-,2^+,3^+,4^+) = 
\frac{i\left(
s_{12
}^2-2s_{24
}^2
\right)[32]
}{\la23\ra\la24\ra\la34\ra[21][31]}.
\end{equation}
Following the same procedure for the other channels and helicity configurations quickly
yields:
\begin{align}
	C_{3;23}^{[2]}(1^-,2^+,3^+,4^+) &=
-\frac{i\la12\ra[32]^2
}{\la24\ra^2[21]}
	, \\
	C_{3;34}^{[2]}(1^-,2^+,3^+,4^+) &=
-\frac{is_{12}[32][43]
}{\la23\ra\la34\ra[31]^2}
	, \\
	C_{3;41}^{[2]}(1^-,2^+,3^+,4^+) &=
-\frac{i\left(
s_{23
}^2-2s_{24
}^2
\right)\la12\ra
}{\la23\ra\la24\ra^2\la34\ra[41]},
\end{align}
and for the $-+-+$ amplitude:
\begin{align}
	C_{3;12}^{[2]}(1^-,2^+,3^-,4^+) &=
\frac{2is_{12}s_{24}
}{\la24\ra^2[31]^2}, \\
	C_{3;23}^{[2]}(1^-,2^+,3^-,4^+) &=
-\frac{2is_{23}\la13\ra\la23\ra
}{\la24\ra^3[41]}, \\
	C_{3;34}^{[2]}(1^-,2^+,3^-,4^+) &= C_{3;12}^{[2]}(1^-,2^+,3^-,4^+), \\
	C_{3;41}^{[2]}(1^-,2^+,3^-,4^+) &= C_{3;23}^{[2]}(1^-,2^+,3^-,4^+),
\end{align}
with all other coefficients vanishing.

For the bubble coefficients, we minimise the number of triangle subtractions by choosing $\chi=p_1$ for $K_1=p_1+p_2$ and $\chi=p_2$
for $K_1=p_2+p_3$. Again, this has the benefit of giving us results directly in terms of the external
momenta and yielding simple analytic forms. The the individual solutions for $y_+$ and
$y_-$ in the triangle subtractions can lead to some spurious denominators and square roots. It is,
therefore, beneficial to perform
the sum over the two solutions algebraically. By considering the boundary behaviour at large $t$ we
can write the solution for $y_\pm$ as:
\begin{align}
	y_\pm &= \alpha_{1,\pm} t + \alpha_{2,\pm} + \frac{1}{t}\alpha_{3,\pm} +
	\mathcal{O}\left(\frac{1}{t^2}\right)\\
	\alpha_{1,\pm} &= \frac{\gammab \AB{\kf1}{K_3}{\kf1}-S_1\AB{\chi}{K_3}{\chi}\pm\sqrt{\alpha}}
	{2S_1\AB{\chi}{K_3}{\kf1}}\\
	\alpha_{2,\pm} &= \frac{1}{2}\left( 1\mp\frac{\sqrt{\alpha}\gammab(S_1+S_2)}{\alpha} \right)
	\\
	\alpha_{3,\pm} &= \pm\frac{1}{4}\left(
	\frac{\sqrt{\alpha}\AB{\chi}{K_3}{\kf1}(S_1-4\mu^2)}{\alpha}-\frac{\sqrt{\alpha}\gammab^2
	S_1\AB{\chi}{K_3}{\kf1}(S_1+S_2)^2}{\alpha^2}\right)\\
	\alpha &= (\gammab\AB{\kf1}{K_3}{\kf1}+S_1\AB{\chi}{K_3}{\chi})^2-4\gammab^2 S_1S_3
	\label{eq:alpha}
\end{align}
The coefficient is then left as a function of $\alpha_{i,\pm}$ and by using sets of identities such as,
\begin{equation}
	\frac{1}{2}\sum_{\pm}\alpha_{1,\pm}^2 =\frac{
	(\gammab \AB{\kf1}{K_3}{\kf1}-S_1\AB{\chi}{K_3}{\chi})^2+\alpha}{4S_1^2\AB{\chi}{K_3}{\kf1}^2}
\end{equation}
we are able to identify terms free from square roots after the summation is performed. Such a
procedure is systematic and has been performed using symbolic manipulation in FORM
\cite{Vermaseren:FORM}. Therefore, after a little algebra, we quickly arrive at the expressions:
\begin{align}
	C_{2;12}^{[2]}(1^-,2^+,3^+,4^+) &= 
\frac{2i\left(
s_{13}-s_{23}
\right)\la13\ra\la14\ra
}{\la12\ra\la23\ra\la24\ra\la34\ra^2[21]}\\
	C_{2;23}^{[2]}(1^-,2^+,3^+,4^+) &=
-\frac{2is_{24}\left(
s_{24}-s_{12}
\right)\la12\ra[32]^2
}{s_{23
}^3\la24\ra^2[21]}
	\\
	C_{2;23}^{[2]}(1^-,2^-,3^+,4^+) &=
\frac{2i\left(
2s_{12}-3s_{23}
\right)\la12\ra^2[41]
}{3\la14\ra\la23\ra^3[21]^2[32]}\\
	C_{2;12}^{[2]}(1^-,2^+,3^-,4^+) &=
\frac{2i\left(
5s_{12}+2s_{23}
\right)\la13\ra^2
}{3s_{12
}^2\la24\ra^2}\\
	C_{2;23}^{[2]}(1^-,2^+,3^-,4^+) &=
\frac{2i\left(
2s_{12}+5s_{23}
\right)\la13\ra^2
}{3s_{23
}^2\la24\ra^2},
\end{align}
with all other bubble coefficients evaluating to zero.

The resulting expressions for $R_4^g$ agree numerically with the known analytic results
\cite{Bern:massuni}. The form of
the rational term is slightly different, and in some cases less compact, than those obtained through
on-shell recursion but include both direct recursive and cut completion terms. Here we have
the advantage that we do not need to calculate any completion terms for the cut-constructible parts
and there are no problems associated with the factorisation in complex momenta.

\subsection{The five- and six-point all-plus amplitudes}

In this section we demonstrate how the technique applies to higher point amplitudes in a
straightforward way. The finite helicity configuration with all gluons carrying positive helicity is
a particularly simple example since there are no triangle or bubble type contributions. We only
need a single five-point tree amplitude to evaluate these all-plus configurations up to the six-point
level:
\begin{align}
	A_5^{(0)}(1_s,2^+,3^+,4^+,5_s) = \frac{i\mu^2\BB{2}{1(2+3)}{4}}{\A 23\A 34\AB 212\AB 454}
\end{align}

We can again make use of eq. \eqref{eq:clim} to quickly determine the leading $\mu^2$ dependence of the
quadruple cut which results in:
\begin{align}
	R_5^g(1^+,2^+,3^+,4^+,5^+) &= -\frac{1}{6}C^{[4]}_{4;12}(1^+,2^+,3^+,4^+,5^+) + \text{cyclic perms.} \\
	C^{[4]}_{4;12}(1^+,2^+,3^+,4^+,5^+) &=
\frac{2i[21][43][53][54]
}{\la12\ra\text{tr}_5(4,1,5,3)}
	\label{eq:A5ppppp}
\end{align}
where 
\begin{equation}
	\text{tr}_5(1,2,3,4) = \AB{1}{234}{1} -\AB{1}{432}{1}
	\label{eq:tr5def}
\end{equation}
For the six point amplitude we have three independent coefficients which also follow from a similar
procedure:
\begin{align}
	&R_6^g(1^+,2^+,3^+,4^+,5^+,6^+) = \nonumber\\ 
	&\hspace{2cm}-\frac{1}{6}C^{[4]}_{4;123}(1^+,2^+,3^+,4^+,5^+,6^+)
	-\frac{1}{6}C^{[4]}_{4;12|34}(1^+,2^+,3^+,4^+,5^+,6^+)\nonumber\\
	&\hspace{2cm}-\frac{1}{12}C^{[4]}_{4;12|45}(1^+,2^+,3^+,4^+,5^+,6^+)
	+ \text{cyclic perms.} \\
	&C^{[4]}_{4;123}(1^+,2^+,3^+,4^+,5^+,6^+) =
\frac{2i\left(
s_{45}\AB{6}{1+2}{3}[51][64]^2-s_{46}\AB{5}{1+2}{3}[54]^2[61]
\right)[56]
}{\la12\ra\la23\ra\text{tr}_5(5,4,6,1)\text{tr}_5(5,4,6,3)}
	\\
	&C^{[4]}_{4;12|34}(1^+,2^+,3^+,4^+,5^+,6^+) =
\frac{2i\AB{5}{1+2}{6}\AB{6}{1+2}{5}[12][43][65]^2
}{\la12\ra\la34\ra\text{tr}_5(5,2,6,1)\text{tr}_5(5,4,6,3)}\\
	&C^{[4]}_{4;12|45}(1^+,2^+,3^+,4^+,5^+,6^+) =
\frac{2i\left(
\AB{3}{1+2}{3}\AB{6}{1+2}{6}-s_{36}s_{12}
\right)[12][54][63]^2
}{\la12\ra\la45\ra\text{tr}_5(2,3,6,1)\text{tr}_5(5,3,6,4)}
	\label{eq:A6pppppp}
\end{align}
These expressions all agree with the previous analytic expressions \cite{Bern:massuni,Bern:1lrec1}.

\subsection{The five-point MHV amplitude}

As a more involved example we present analytic expressions for the five-point MHV configuration
$R_5^g(1^+,2^+,3^+,4^-,5^-)$. This amplitude has been derived previously using a string based
analysis \cite{Bern:5g} and more recently using on-shell recursion relations \cite{Bern:bootstrap}
\begin{align}
	R_5^g(1^+,2^+&,3^+,4^-,5^-) = \sum_{k=1}^{5}
	\bigg(-\frac{1}{6}C^{[4]}_{4;k(k+1)}(1^+,2^+,3^+,4^-,5^-)
	-\frac{1}{2}C^{[2]}_{3;k(k+1)(k+2)}(1^+,2^+,3^+,4^-,5^-)\nonumber\\
	-&\frac{1}{2}C^{[2]}_{3;k(k+1)|(k+2)(k+3)}(1^+,2^+,3^+,4^-,5^-)
	-\frac{s_{k(k+1)}}{6}C^{[2]}_{2;k(k+1)}(1^+,2^+,3^+,4^-,5^-)\bigg)
	\label{eq:R5MHVbasis}\\
	&\hspace{-1cm}=-\frac{1}{6}C^{[4]}_{4;{\rm total}}-\frac{1}{2}C^{[2]}_{3;{\rm total}}
	-\sum_{k=1}^5 \frac{s_{k(k+1)}}{6}C^{[2]}_{2;k(k+1)}(1^+,2^+,3^+,4^-,5^-)
\end{align}
The computation of box contributions follows in the same way as for the all-plus configuration
though the lack of symmetry means each coefficient must be computed separately. The additional tree
level amplitude required is given by \cite{Badger:mrec,Forde:mscalar},\footnote{We note that one must
compensate for the different normalisation of the spinor products used in \cite{Badger:mrec}.}
\begin{align}
	A_5^{(0)}(1_S,2^-,3^+,4^+,5_S) = 
	\frac{i\AB{2}{1(3+4)5}{4}^2}{\A 23\A 34\AB 212\AB 454\BB{2}{(3+4)5}{4}}
	-\frac{i\mu^2\B 34^3}{s_{234}\B 23 \BB{2}{(3+4)5}{4}}.
\end{align}
The sum of all five box contributions then yields the following simple result:
\begin{align}
	&C^{[4]}_{4;{\rm total}} = \sum_{k=1}^5 C^{[4]}_{4;k(k+1)}(1^+,2^+,3^+,4^-,5^-) =
	\frac{2i}{\text{tr}_5(1,2,3,4)}\bigg(\nonumber\\&
\frac{\la14\ra^2\la25\ra^2[53]^2[21]^3}{\la12\ra^2\la34\ra[43][51][52]}
-\frac{2\la14\ra\la25\ra\la45\ra[31][53][21]^2}{\la12\ra\la34\ra[43][51]}
+\frac{\la45\ra^2[31]^2[52][21]}{\la34\ra[43][51]}
+\frac{\la45\ra[31][32][21]}{[54]}\nonumber\\&
+\frac{\la45\ra^2[43][53][21]}{\la12\ra[54]}
-\frac{2\la24\ra\la35\ra\la45\ra[31][32]^2[41]}{\la15\ra\la23\ra[43][51]}
+\frac{\la45\ra^2[31]^2[32][42]}{\la15\ra[43][51]}\nonumber\\&
+\frac{\la24\ra^2\la35\ra^2[32]^3[41]^2}{\la15\ra\la23\ra^2[42][43][51]}
+\frac{\la45\ra^2[32][41][51]}{\la23\ra[54]}
	\bigg),
	\label{eq:c4tot}
\end{align}
while for the sum of the ten triangles we have,
\begin{align}
	&C^{[2]}_{3;{\rm total}} =
-\frac{2i\la24\ra[32]^2[21]^3}{\AB{2}{5+1}{2}^2[42][51][52]}
-\frac{i\la23\ra[32]^3[21]^3}{\AB{2}{5+1}{2}^2[42]^2[51][52]}
-\frac{i[43][21]^3}{\la23\ra[42]^2[51][54]i}\nonumber\\&
-\frac{i\la12\ra[32]^3[21]^3}{\AB{2}{5+1}{2}^2[42][43][52]^2}
-\frac{2i\la24\ra^2\la25\ra[32][21]^2}{\la12\ra\la23\ra\AB{2}{5+1}{2}^2[51]}
-\frac{2i\la25\ra[32]^3[21]^2}{\AB{2}{5+1}{2}^2[42][43][52]}\nonumber\\&
-\frac{i\la15\ra[53]^3[21]^2}{\la12\ra^2[43][51][52]^2[54]}
+\frac{2i[31][21]^2}{\la23\ra[42][51][54]}
-\frac{i\la24\ra^2\la25\ra^2[32][52][21]}{\la12\ra^2\la23\ra\AB{2}{5+1}{2}^2[51]}\nonumber\\&
-\frac{2i\la24\ra\la25\ra^2[32]^2[21]}{\la12\ra\la23\ra\AB{2}{5+1}{2}^2[43]}
-\frac{i\la24\ra^2\la25\ra^2[32][42][21]}{\la12\ra\la23\ra^2\AB{2}{5+1}{2}^2[43]}
+\frac{i\la45\ra[53]^2[21]}{\la12\ra^2[51][52][54]}\nonumber\\&
-\frac{i\la15\ra[32][53]^2[21]}{\la12\ra^2[43][52]^2[54]}
+\frac{i[31]^2[21]}{\la23\ra[43][51][54]}
+\frac{i\la45\ra[32][41]^2}{\la23\ra^2[42][43][54]}
+\frac{i[31]^2[32]}{\la12\ra[43][51][54]}\nonumber\\&
-\frac{i\la34\ra[31][32][41]^2}{\la23\ra^2[42][43][51][54]}
+\frac{2i[31][32]^2}{\la12\ra[43][52][54]}
-\frac{i[32]^3[51]}{\la12\ra[43][52]^2[54]}.
\end{align}
The bubble coefficients are more complicated for higher point amplitudes due to the
appearance of additional poles in the triangle subtraction terms. In order to give compact expressions it
is convenient to leave the sum over $y_\pm$ unexpanded. We choose a value of $\chi=p_k$ for the
$K_1=p_k+p_{k+1}$ channel which ensures there are a maximum of two triangle subtraction terms.
Furthermore it is useful to note that to evaluate the coefficient at the boundary of the $\mu$
contour it is sufficient to use the following values for the non-vanishing integrals:
\begin{align}
	Y_0&=1 & Y_1&=0 & Y_3&=-\frac{\mu^2}{3S_1} \\
	T_1&=-\frac{S_1\AB{\chi}{K_3}{\kf1}}{2\gammab\Delta} & T_2 &= 0 & T_3 &= -\frac{\mu^2S_1\AB{\chi}{K_3}{\kf1}^3}{3\gammab^3\Delta^2}
	\label{}
\end{align}
The justification for this can be seen be examining the expanded forms of the $\inf$ operations
given in appendix \ref{app:mucount}.
\begin{align}
	&C^{[2]}_{2;12}(1^+,2^+,3^+,4^-,5^-) =
	\frac{2i([32][51]-3[31][52])[32]^2}{\la12\ra^2[21][43][52]^2[54]}\nonumber\\&
	-2i\sum_{\sigma=\pm}
\frac{
\la15\ra(\la24\ra-\alpha^{1;p_2}_{1,\sigma}\la14\ra)^2(\alpha^{1;p_2}_{1,\sigma}\la15\ra-\la25\ra)[21]([31]+\alpha^{1;p_2}_{1,\sigma}[32])^2[52]
}{
D_{33}(\alpha^{1;p_2}_{1,\sigma},1,1+2)\la12\ra^2\la34\ra\AB{5}{1+2}{5}^2[43][51]
},\\
	&C^{[2]}_{2;23}(1^+,2^+,3^+,4^-,5^-) = 
\frac{2i([31][42]-3[21][43])[31]^2}{\la23\ra^2[32][43]^2[51][54]}\nonumber\\&
-2i\sum_{\sigma=\pm}
\frac{(\alpha^{2;p_{51}}_{1,\sigma}\la24\ra-\la34\ra)(\la35\ra-\alpha^{2;p_{51}}_{1,\sigma}\la25\ra)^2\AB{2}{5+1}{3}([21]+\alpha^{2;p_{51}}_{1,\sigma}[31])^2[32]
}{
D_{55}(\alpha^{2;p_{51}}_{1,\sigma},2,2+3)\la15\ra\la23\ra^2\AB{4}{5+1}{4}^2[42][51]},
\end{align}
\begin{align}
	&C^{[2]}_{2;34}(1^+,2^+,3^+,4^-,5^-) =
	\frac{2i\la24\ra\la25\ra^2\la35\ra}{3\la12\ra\la15\ra\la23\ra^4[43]}
+\frac{10i\la25\ra^2\la45\ra}{3\la12\ra\la15\ra\la23\ra^3[43]}
-\frac{4i\la24\ra[21][41]^2}{\la23\ra^3[42][43][51][54]}\nonumber\\&
-\frac{4i[21][31][41]}{\la23\ra^2[42][43][51][54]}
+\frac{2i[21][32][41]^2}{\la23\ra^2[42]^2[43][51][54]}
+\frac{2i[21][41]^2[53]}{\la23\ra^2[42][43][51][54]^2}\nonumber\\&
-2i\sum_{\sigma=\pm}\bigg(
\frac{(\alpha^{3;p_{12}}_{1,\sigma}\la35\ra-\la45\ra)\AB{3}{1+2}{4}[21][43](\alpha^{3;p_{12}}_{1,\sigma})^4
}{D_{11}(\alpha^{3;p_{12}}_{1,\sigma},3,3+4)\la12\ra\AB{5}{1+2}{5}^2[53]}
+\frac{\alpha^{3;p_2}_{1,\sigma}D_{51}(\alpha^{3,p_2}_{1,\sigma},3,3+4)^2\AB{3}{4}{2}[42](\alpha^{3;p_2}_{1,\sigma})^3}
{D_{55}(\alpha^{3;p_2}_{1,\sigma},3,3+4)\la15\ra\AB{2}{3+4}{2}^2[51]}\nonumber\\&
+\frac{8D_{51}(\alpha^{3;p_2}_{1,\sigma},3,3+4)^2\la23\ra^2\la34\ra[42]^3([32]+\alpha^{3;p_2}_{1,\sigma}[42])(\alpha^{3;p_2}_{1,\sigma})^3}
{3D_{55}(\alpha^{3;p_2}_{1,\sigma},3,3+4)s_{34}\la15\ra\AB{2}{3+4}{2}^4[51]}\nonumber\\&
+\frac{2D_{51}(\alpha^{3;p_2}_{1,\sigma},3,3+4)D'_{51}(\alpha^{3;p_2}_{1,\sigma},\alpha^{3,p_2}_{1,\sigma},3,3+4)\la34\ra[42]([32]+\alpha^{3;p_2}_{1,\sigma}[42])(\alpha^{3;p_2}_{1,\sigma})^3}
{D_{55}(\alpha^{3;p_2}_{1,\sigma},3,3+4)\la15\ra\AB{2}{3+4}{2}^2[51]}\nonumber\\&
-\frac{D_{51}(\alpha^{3;p_2}_{1,\sigma},3,3+4)^2D'_{55}(\alpha^{3;p_2}_{1,\sigma},\alpha^{3;p_2}_{1,\sigma},3,3+4)\la34\ra[42]([32]+\alpha^{3;p_2}_{1,\sigma}[42])(\alpha^{3;p_2}_{1,\sigma})^3}
{D_{55}(\alpha^{3;p_2}_{1,\sigma},3,3+4)^2\la15\ra\AB{2}{3+4}{2}^2[51]}\nonumber\\&
-\frac{\alpha^{3;p_2}_{1,\sigma}D_{51}(\alpha^{3;p_2}_{1,\sigma},3,3+4)^2s_{34}[42]([32]+\alpha^{3,p_2}_{1,\sigma}[42])(\alpha^{3;p_2}_{1,\sigma})^2}
{D_{55}(\alpha^{3;p_2}_{1,\sigma},3,3+4)\la15\ra\AB{2}{3+4}{2}^2[43][51]}\nonumber\\&
+\frac{2D_{51}(\alpha^{3;p_2}_{1,\sigma},3,3+4)^2\left(
2\alpha^{3;p_2}_{1,\sigma}\alpha^{3,p_2}_{1,\sigma}s_{34}+1
\right)[42]([32]+\alpha^{3,p_2}_{1,\sigma}[42])\alpha^{3,p_2}_{1,\sigma}}{D_{55}(\alpha^{3;p_2}_{1,\sigma},3,3+4)\la15\ra\AB{2}{3+4}{2}^2[43][51]}
\bigg),
	\\
	&C^{[2]}_{2;45}(1^+,2^+,3^+,4^-,5^-) = 0, \\
	&C^{[2]}_{2;51}(1^+,2^+,3^+,4^-,5^-) = C^{[2]}_{2;34}(3^+,2^+,1^+,5^-,4^-),
\end{align}
where we define the following functions,
\begin{align}
	\alpha^{k;K_3}_{1,\pm} &= \frac{\AB{k}{k+1}{k}\AB{k+1}{K_3}{k+1}-s_{k(k+1)}\AB{k}{K_3}{k}\mp\alpha}{2s_{k(k+1)}\AB{k}{K_3}{k+1}},\\
	\alpha^{k;K_3}_{3,\pm} &= \mp\frac{\sqrt{\alpha}}{\alpha}\AB{k}{K_3}{k+1},\\
	D_{xy}(\alpha,p,q) &= -\alpha^2\A{x}{p}\B{\fl q}{y}
	+\alpha\left( \AB x{\fl q}y -\frac{q^2}{\AB pqp }\AB xpy \right)
	+\A{x}{\fl q}\B{p}{y},\\
	D'_{xy}(\alpha,\beta,p,q) &= -\left(\frac{1}{2q\cdot p}+2\alpha\beta\right)\A xp \B{\fl
	q}{y}+\beta\left( \AB{x}{\fl q}{y}-\frac{q^2}{2q \cdot p}\AB{x}{p}{y} \right),\\
	\label{eq:tsfuncs}
	\text{where }\fl q &= q - \frac{q^2}{2p\cdot q}p. \nonumber
\end{align}
The function $\alpha$ is given in equation \eqref{eq:alpha}. These expressions have been checked
numerically against the known results of \cite{Bern:5g}.

\subsection{The six-point $-+-+-+$ amplitude}

As a final analytic example we consider the NMHV six-point amplitude with alternating helicities,
$R^g_6(1^-,2^+,3^-,4^+,5^-,6^+)$. This amplitude has been considered in previous analytic studies
based on Feynman diagrams \cite{Xiao:6ptrat} as well as more recent numerical unitarity approaches
\cite{Berger:blackhat,Giele:dduni}. Here the analytic expression can be reduced to a set of seven
independent coefficients through the symmetries of the external helicities. We first define three
transformation identities:
\begin{align}
	\alpha_l: (1,2,3,4,5,6) &\mapsto (l+1,l+2,l+3,l+4,l+5,l+6) \\
	\alpha_l^\dagger: (1,2,3,4,5,6) &\mapsto (l+1,l+2,l+3,l+4,l+5,l+6) |_{\la\ra\leftrightarrow
	[]}\\
	\beta: (1,2,3,4,5,6) &\mapsto (6,5,4,3,2,1)
\end{align}
The rational contribution to this amplitude can then be written as,
\begin{align}
	&R_6^g(1^-,2^+,3^-,4^+,5^-,6^+) = 
	-\frac{1}{6}\sum_{k=0,2,4} 
	\Bigg(
	 \alpha_k\left(C_{4;123}^{[4]}\right)
	+\alpha_k\left(C_{4;12|34}^{[4]}\right)\nonumber\\&
	+\alpha_k\left(C_{4;12|45}^{[4]}\right)
	+\alpha_{k+1}^\dagger\left(C_{4;123}^{[4]}\right)
	+\alpha_{k+1}^\dagger\left(C_{4;12|34}^{[4]}\right)
	+\alpha_{k+1}^\dagger\left(C_{4;12|45}^{[4]}\right)
	\nonumber\\&
	 3\alpha_k\left(C_{3;1234}^{[2]}\right)
	+3\alpha_k\left(C_{3;123x45}^{[2]}\right)
	+3\alpha_{k+1}^\dagger\left(C_{3;123x45}^{[2]}\right)
	+3\alpha_{k+5}\left(\beta\left(C_{3;123x45}^{[2]}\right)\right)\nonumber\\&
	+3\alpha_{k}^\dagger\left(\beta\left(C_{3;123x45}^{[2]}\right)\right)
	+\alpha_0\left(C_{3;12x34x56}^{[2]}\right)
	+\alpha_1^\dagger\left(C_{3;12x34x56}^{[2]}\right)\nonumber\\&
	+s_{k+1,k+2}\alpha_k\left(C_{2;12}^{[2]}\right)
	+s_{k+2,k+3}\alpha_{k+5}\left(\beta\left(C_{2;12}^{[2]}\right)\right)
	+s_{k+1,k+2,k+3}\alpha_k\left(C_{2;123}^{[2]}\right)
	\Bigg).
\end{align}
The procedure for computing analytic forms for these coefficients has been automated with the help
of symbolic manipulations in FORM \cite{Vermaseren:FORM}. In particular the procedure does not
require any prior algebraic manipulation of the tree level amplitude although the final form of the
coefficient will depend on the form of the integrand. We have found it useful to choose a basis for
the loop momenta that reflects the symmetry of the coefficient. Therefore adjacent massless legs
were chosen where possible. However for the 2-mass easy box configuration it is much simpler to
choose the two, non-adjacent, massless legs. The solution to the on-shell constraints in this case
can be easily obtained from \eqref{eq:csol} using an exchange of $K_2\leftrightarrow K_3$.
\begin{align}
	&C_{4;123}^{[4]}(1^-,2^+,3^-,4^+,5^-,6^+) = 
	\frac{2i}{\text{tr}_5(\eta_{1,1+2|3}^a,5,4,6)}\bigg(\nonumber\\&\hspace*{1cm}
\frac{\text{tr}_5(\eta_{1,2},5,4,6)^2\text{tr}_5(\eta_{3,2},5,4,6)^2[64]}{\AB{4}{6}{5}\AB{6}{4}{5}[21][32]\text{tr}_5(5,1,6,4)\text{tr}_5(5,3,6,4)}
-\frac{\la13\ra^4\la5|46|5\ra[64]}{s_{123}\la12\ra\la23\ra\la46\ra}\bigg)\\
	&C_{4;12|34}^{[4]}(1^-,2^+,3^-,4^+,5^-,6^+) =
	\frac{
	2i\,\text{tr}_5(\eta_{1,2},5,1+2,6)^2\text{tr}_5(\eta_{3,4},5,1+2,6)^2
	}{
	s_{12}s_{34}s_{56}\AB{6}{1+2}{5}^2\text{tr}_5(5,2,6,1)\text{tr}_5(5,4,6,3)
	}\\&
	C_{4;12|45}^{[4]}(1^-,2^+,3^-,4^+,5^-,6^+) = C_{4;12|34}^{[4]}(1^-,2^+,5^-,4^+,3^-,6^+)
\end{align}
where we have defined complex momenta $\eta_{i,j}^\mu=\tfrac{1}{2}\AB{i}{\gamma^\mu}{j}$ and
$\eta_{i,j|k}^{a,\mu}=\tfrac{1}{2}\AA{i}{\gamma^\mu \slashed{j}}{k}$.

Following a similar automated procedure expressions for the triangle and bubble contributions have
also been generated and have been checked numerically against the known results. 
However, since the expressions for these remaining coefficients are quite lengthy and
not particularly illuminating, we include a {\tt Mathematica} input file, {\tt R6mpmpmpCoeffs.m}, containing the coefficients
which follows the notation of the {\tt S@M} package \cite{Maitre:SAM}. To present more compact forms
ofthe triangle subtractions terms it
was necessary to use a slightly expanded basis functions than was used in the five-point example:
\begin{align}
	D^{a}_{xyz}(\alpha,p,q) &= -\alpha^2\A{x}{p}\BA{\fl q}{y}{z}
	+\alpha\left( \AA x{\fl qy}z -\frac{q^2\AA x{py}z}{\AB pqp } \right)
	+\A{x}{\fl q}\BA{p}{y}{z},\\
	D^{b}_{xyz}(\alpha,p,q) &= -\alpha^2\B{x}{\fl q}\AB{p}{y}{z}
	+\alpha\left( \BB x{\fl qy}z -\frac{q^2\BB x{py}z}{\AB pqp } \right)
	+\B{x}{p}\AB{\fl q}{y}{z},\\
	D^{a'}_{xyz}(\alpha,\beta,p,q) &= 
	-\left(\frac{1}{2q\cdot p}+2\alpha\beta\right)\A xp \BA{\fl q}{y}{z}
	+\beta\left( \AA{x}{\fl qy}{z}-\frac{q^2\AA{x}{py}{z}}{2q \cdot p} \right),\\
	D^{b'}_{xyz}(\alpha,\beta,p,q) &= 
	-\left(\frac{1}{2q\cdot p}+2\alpha\beta\right)\A x{\fl q} \AB p{y}{z}
	+\beta\left( \BB{x}{\fl qy}{z}-\frac{q^2\BB{x}{py}{z}}{2q \cdot p} \right),
\end{align}
with all other definitions given in equation \eqref{eq:tsfuncs}. The bubble coefficients are then
given by:
\begin{align}
	C_{2;12}^{[2]} &= -i C_{2;12}^{[2],{\rm bub}}-\frac{1}{2}\sum_{\sigma=\pm} 
	C_{2;12}^{[2], K_2=p_6}+C_{2;12}^{[2], K_2=p_5+p_6}+C_{2;12}^{[2],
	K_2=p_4+p_5+p_6}\\
	C_{2;123}^{[2]}(\kf1) &= -i C_{2;123}^{[2],{\rm bub}}(\kf1)\nonumber\\&-\frac{1}{2}\sum_{\sigma=\pm} 
	C_{2;123}^{[2], K_2=p_6}(\kf1)+C_{2;123}^{[2],K_2=p_5+p_6}(\kf1)+C_{2;123}^{[2],
	K_2=p_1+p_2}(-\kf 1)
\end{align}
where $\chi=p_1$ has been chosen for the basis in both cases.

Since the procedure for analytic extraction has been automated it has also been possible to generate
unsimplified expressions for all other configurations with up to six gluons which have been checked
numerically against existing results. These expressions are available from the author on request.

\subsection{Remaining helicity configurations}

Since the extraction is purely algebraic is also well suited numerical extraction.
One possible way to do this is to use the discrete Fourier projections to find the value of
$C_4^{[4]}$. From eq. \eqref{eq:ddbox} the box contribution can then be written as,
\begin{align}
	C_4^{[4]} &= \frac{i}{2p_\mu}\sum_{\sigma=\pm}\sum_{k=0}^{p_\mu-1}
	\frac{1}{\mu_k^4}A_1 A_2 A_3 A_4(\bl_1^{\sigma},\mu_k), 
	\label{eq:FPbox}
	\\
	\mu_k&=\mu_\infty \exp\left(2\pi i\tfrac{k}{p_\mu}\right),\nonumber
\end{align}
where $\mu_\infty$ is some large constant and $p_\mu+1$ is the number of points around the circle
of integration, on which the residue is evaluated. It is straightforward to write down equivalent
expressions for the triangle,
\begin{align}
	&\Rightarrow C_3^{[2]} = 
	-\frac{1}{2p_\mu p_t}
	\sum_{\sigma}\sum_{k=0}^{p_\mu-1}\sum_{l=0}^{p_t-1}\frac{1}{\mu_k^2}A_1 A_2 A_3
	(\bl_1^\sigma,t_l,\mu^2_k),\\
	&\mu_k=\mu_\infty \exp\left(2\pi i\tfrac{k}{p_\mu}\right), \qquad\qquad
	t_l=t_\infty \exp\left(2\pi i\tfrac{l}{p_t}\right),
	\label{eq:ddtri}
\end{align}
and bubble terms,
\begin{align}
	C_2^{ {\rm bub} [2]} &= 
	-\frac{i}{p_\mu p_tp_y}\sum_{k=0}^{p_\mu-1}\sum_{l=0}^{p_t-1}\sum_{q=0}^{p_y-1}
	\sum_{s=0}^{2}\frac{Y_s}{\mu^2_ky_q^s} A_1A_2(\bl_1(y_q,t_{l;-1},\mu^2_k))\\
	C_2^{ {\rm tri}(K_3) [2]} &= 
	-\frac{1}{2p_\mu p_t}\sum_{\sigma=\pm}\sum_{k=0}^{p_\mu-1}\sum_{l=0}^{p_t-1}
	\sum_{s=1}^{3}\frac{T_s}{\mu^2_kt_{l;1}^s} A_1A_2A_3(\bl_1(y_\sigma,t_{l;1},\mu^2_k))\\
	\mu_k&=\mu_\infty \exp\left(2\pi i\tfrac{k}{p_\mu}\right) \qquad
	t_{l;\alpha}=t_\infty^\alpha \exp\left(2\pi i\tfrac{l}{p_t}\right) \qquad
	y_q=y_\infty \exp\left(2\pi i\tfrac{q}{p_y}\right).
	\nonumber
\end{align}
Since the locations of the boundaries depend upon the order in which we take $t$ and $y$ large, we must
take $\alpha=-1$ for the bubble terms and $\alpha=1$ in the triangle subtraction terms.

The method of Fourier projections has already been shown to be extremely efficient in both the
\texttt{BlackHat} code \cite{Berger:blackhat} and with the OPP technique \cite{Mastrolia:OPPopt}. In
general one can arbitrarily increase the numerical
accuracy by increasing the size of the circle, $\mu_\infty$, and the number of points $p_\mu\geq
4$. We note that, in order to find a stable numerical implementation at fixed precision, it may well be
necessary to make explicit subtractions of the poles from pentagon integrals
\cite{Ossola:rational,Giele:dduni,Britto:ddmassless}. This could be done in an analogous way to
the treatment of the cut-constructible contributions in the \texttt{BlackHat} code by computing
values of $\mu^2$ which put the additional propagator on-shell
\cite{Berger:blackhat}.

We have tested our method for the direct extraction of the rational terms for the remaining one-loop
gluon helicity amplitudes with up to six external legs using a numerical approach with discrete Fourier
transforms. In order to compare with the known results it was sufficient to proceed without the
subtraction of higher order residues in place of considering a large radius for the integration
contour. The required tree level amplitudes have been generated using on-shell recursion relations 
literature \cite{Badger:mrec,Forde:mscalar}. The rational terms have been checked against the known analytic results
\cite{Bern:bootstrap,Berger:1lmhv,Berger:genhels} and the
numerical results of references \cite{Ellis:6ptnum,Giele:dduni,Giele:rocket}. 
We have also checked these results against expressions generated analytically through an automated
implementation of the direct extraction method.
In order to achieve
numerical precision of at least $10^{-6}$ it was necessary to use a radius for each contour integral of
$10^4$ (i.e. the parameters $\mu_\infty,t_\infty,y_\infty$) and in some cases evaluate at a
large number of points around the circle (i.e. $p_\mu,p_t,p_y\sim10$). This was achieved using symbolic
manipulation in FORM \cite{Vermaseren:FORM} and numerical procedures in Maple. In order to obtain a
more stable implementation,
subtraction of the residues of high point functions from the complex plane
\cite{Berger:blackhat,Ossola:OPP} would allow setting all the radii to 1 and minimise the number of
points required around the circle. It is expected that these improvements would considerably improve
the speed of the algorithm but we leave such an implementation for future study.

\section{Rational contributions to quark amplitudes \label{sec:quarks}}

\begin{figure}[t]
	\begin{center}
		\psfrag{1}{$1_q$}
		\psfrag{2}{$2$}
		\psfrag{3}{$3$}
		\psfrag{4}{$4_q$}
		\includegraphics[width=12cm]{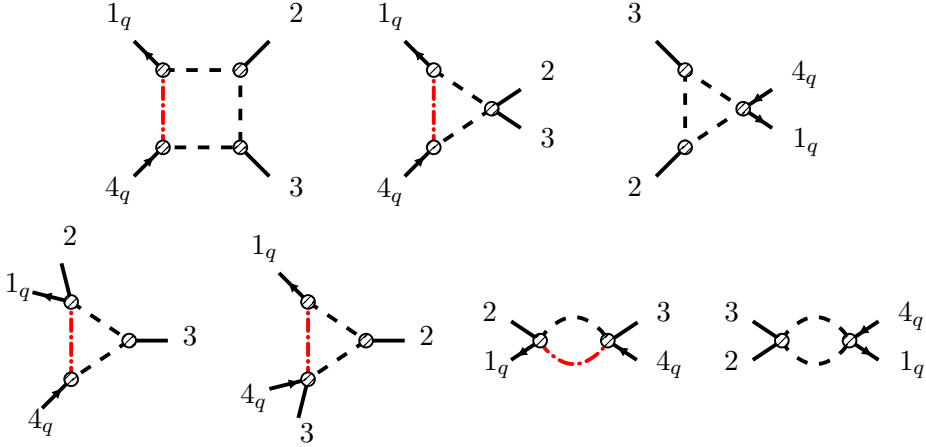}
	\end{center}
	\caption{Integral basis for the rational part of the leading colour contribution to the
	$gg\to q\bar{q}$ amplitude, $R_4^{[L]}$.}
	\label{fig:qqbasis}
\end{figure}

As an example of how the method can apply equally well to amplitudes with external fermions we
re-compute the rational parts of the four-point process with a pair of massless quarks, $gg\to
q\bar{q}$. The $n$-point colour ordered amplitude can be written as:
\begin{align}
	A_n^f &= A_n^{[L]} + \frac{1}{N_c^2} A_n^{[R]} + \frac{N_f}{N_c} A_n^{L,[1/2]} +
	\frac{N_s}{N_c} A_n^{[0]}
	\\
	&= 
	\left(1+\frac{1}{N_c^2}\right) A_n^{[L]} 
	- \frac{1}{N_c^2} A_n^{[SUSY]} \nonumber\\
	&- \left(\frac{N_f}{N_c}+\frac{1}{N_c^2}\right) (A_n^{[L,1/2]}+A_n^{[s]})
	+ \left(\frac{N_s-N_f}{N_c}-\frac{1}{N_c^2}\right) A_n^{[s]}.
	\label{eq:1lqqcolour}
\end{align}
The supersymmetric decomposition ensures that the only rational contributions come from the scalar
term, $A_n^{[s]}$ (just as in the gluon case), and $A_n^{[L]}$ where the fermion line follows the
shortest path through the loop (see reference \cite{Bern:qqggg} for the definitions of the left and right moving primitive amplitudes). 
\begin{equation}
	R_n^f = \left(1+\frac{1}{N_c^2}\right)R_n^{[L]} + \left(\frac{N_s-N_f}{N_c}-\frac{1}{N_c^2}\right) R_n^{[s]}
	\label{eq:qqrat}
\end{equation}
Figure \ref{fig:qqbasis} shows the rational part of $A_4^{[L]}$ written in
terms of the massive integral basis.

In the following calculation we show that it is possible to reconstruct the rational contributions
to the left moving primitive amplitude by looking only at the transverse polarisation of the
$D$-dimensional gluon. This can also be interpreted as introducing a massive scalar in the loop in an
analogous way to the gluon examples. We then simply interpret the $D$-dimensional internal fermion as
four-dimensional massive fermion \footnote{We do not present a formal proof of this fact since in
general there is an ambiguity in the definition of the higher dimensional Clifford algebra. The
definition is scheme dependent but while working within the FDH scheme we simply note that there is no
problem in treating the $D$-dimensional fermion as a massive one for the example presented
here. Indeed $D$-dimensional massive fermions have been treated successfully within the FDH scheme
in a recent paper \cite{Ellis:massdduni}}.

We now construct the tree-level structures using both massive scalars and massive fermions so that
each cut has a uniform internal mass, $\mu^2$.
\begin{align}
	V_3(1_q,2_S,3_Q) &= i\bar{u}(p_1)v(p_3,\mu) \\
	V_3(1_Q,2_S,3_q) &= i\bar{u}(p_1,\mu)v(p_3)
	\label{eq:scalarFR}
\end{align}
where we can use the following representations for the fermion wave-functions:
\begin{equation}
	\bar{u}_\pm(q,\eta) = \frac{\la \eta\mp|(\slashed{q}+\mu)}{\A{\eta \mp|}{\fl q \pm}}, \qquad
	v_\pm(q,\eta) = \frac{(\slashed{q}-\mu)|\eta\mp\ra}{\A{\fl q \pm|}{\eta \mp}}
	\label{eq:Qwf}
\end{equation}
The relevant four-point amplitudes are given by:
\begin{align}
	A_4(1_q,2^+,3_S,4_Q) &= \frac{i\bar{u}(p_1)v(p_4,\mu)\AB{1}{3}{2}}{\AB 232\A 12}, \\
	A_4(1_q,2^-,3_S,4_Q) &= \frac{i\bar{u}(p_1)v(p_4,\mu)\AB{2}{3}{1}}{\AB 232\B 21}, \\
	A_4(1_Q,2_S,3^+,4_q) &= \frac{i\bar{u}(p_1,\mu)v(p_4)\AB{4}{2}{3}}{\AB 323\A 34}, \\
	A_4(1_Q,2_S,3^-,4_q) &= \frac{i\bar{u}(p_1,\mu)v(p_4)\AB{3}{2}{4}}{\AB 323\B 43}, \\
	A_4(1_q^+,2_S,3_S,4_q^-) &= i\AB 421\left( \frac{1}{s_{41}}+\frac{1}{2\AB 121} \right).
\end{align}

The leading mass components of the seven integral coefficients, shown in figure \ref{fig:qqbasis}, are easily computed with the
techniques used for the gluon amplitudes in section \ref{sec:gluons}. Firstly we note that for the
leading colour primitive amplitude all box contributions are zero
since the maximum power of $\mu$ that can appear in the product of the four three-point vertices is
three. The first non-trivial coefficients are therefore the triangle coefficients which we present
here for the three independent helicity configurations. The $+++-$ configuration is given by,
\begin{align}
	C_{3;12}^{[2]}(1_q^+,2^+,3^+,4_{\bar{q}}^-) &= \frac{i\B12\B23}{2\A23\B24}, \\
	C_{3;23}^{[2]}(1_q^+,2^+,3^+,4_{\bar{q}}^-) &= -\frac{i\B 23^2}{2\A 12 \B 24}, \\
	C_{3;34}^{[2]}(1_q^+,2^+,3^+,4_{\bar{q}}^-) &= \frac{i\A14\B12}{2\A23\A13}+\frac{i\A 14^2\B
	12}{\A12\A34\A13}, \\
	C_{3;41}^{[2]}(1_q^+,2^+,3^+,4_{\bar{q}}^-) &=
-\frac{i\left(s_{14}\la13\ra\la24\ra+\left(s_{14}-4s_{24}\right)\la12\ra\la34\ra\right)[32]
}{2\la12\ra\la13\ra\la14\ra\la23\ra[41]}, \\
	\sum_{k=1}^4 C_{3;k(k+1)}^{[2]} (1_q^+,2^+,3^+,4_{\bar{q}}^-) &=
	-\frac{i\B 31(2s_{13}+s_{23})}{\A 12\A 23\B 41},
\end{align}
the $++--$ configuration by, 
\begin{align}
	C_{3;12}^{[2]}(1_q^+,2^+,3^-,4_{\bar{q}}^-) &= \frac{i\A34^2}{2\A12\A24}, \\
	C_{3;23}^{[2]}(1_q^+,2^+,3^-,4_{\bar{q}}^-) &=
	-\frac{i\A14\A23\A34}{2\A12^2\A24}+\frac{i\A34^2\B21}{\A23\A24\B32}, \\
	C_{3;34}^{[2]}(1_q^+,2^+,3^-,4_{\bar{q}}^-) &= C_{3;12}^{[2]}(1_q^+,2^+,3^-,4_{\bar{q}}^-), \\
	C_{3;41}^{[2]}(1_q^+,2^+,3^-,4_{\bar{q}}^-) &= -\frac{i\A14\A23\B21}{2\A12^2\B31},\\
	\sum_{k=1}^4 C_{3;k(k+1)}^{[2]} (1_q^+,2^+,3^+,4_{\bar{q}}^-) &= 
	-\frac{i s_{13}\A34\B21}{s_{23}\A24\B43},
\end{align}
and finally the $+-+-$ configuration,
\begin{align}
	C_{3;12}^{[2]}(1_q^+,2^-,3^+,4_{\bar{q}}^-) &= \frac{i\A24\B41\B43}{2\A23\B42^2} ,\\
	C_{3;23}^{[2]}(1_q^+,2^-,3^+,4_{\bar{q}}^-) &= 
	-\frac{i [31] \left(s_{24} ([31] [42]+[21] [43])-s_{23} [21] [43]\right)}{2 \la 23\ra  [21]
	[32][42]^2},
	\\
	C_{3;34}^{[2]}(1_q^+,2^-,3^+,4_{\bar{q}}^-) &= C_{3;12}^{[2]}(1_q^+,2^-,3^+,4_{\bar{q}}^-), \\
	C_{3;41}^{[2]}(1_q^+,2^-,3^+,4_{\bar{q}}^-) &= -\frac{i\A14\A23\B31}{2\A13^2\B21}, \\
	\sum_{k=1}^4 C_{3;k(k+1)}^{[2]} (1_q^+,2^+,3^+,4_{\bar{q}}^-) &=
	\frac{i\A12\B31^3}{s_{23}\B24}.
	\label{eq:qqggbtri}
\end{align}
There is only one non-zero bubble coefficient for all the four point fermion amplitudes considered
here. This is the $23$ channel for the $+++-$ helicity configuration for which we choose a basis
defined by $K_1=p_2+p_3$ with $\chi=p_1$ ($\Rightarrow \kf 1=p_3$). This choice ensures that
there is only one triangle subtraction term and the two integrands are:
\begin{align}
	C_{2;23}(1_q^+,2^+,3^+,4_{\bar{q}}^-) &= \frac{2\mu^2\B 23 \AB{4}{l_1}{1}}{\A 23\AB{2}{l_1}{2}}
	\left( \frac{1}{s_{41}}+\frac{1}{2\AB{1}{l_1}{1}} \right)\\
	C_{2;23}^{{\rm tri}; K_3=p_1}(1_q^+,2^+,3^+,4_{\bar{q}}^-) &= \frac{2\mu^2\B 23 \AB{4}{l_1}{1}}{\A 23\AB{2}{l_1}{2}}.
\end{align}
Since $T_0=0$, the triangle subtraction term will vanish after substitution and expansion of
the loop momentum. The second term in the pure bubble is also suppressed by an additional power of
the loop momentum and therefore vanishes after taking the large $y$ limit. The single remaining term
is then gives the final value for the coefficient:
\begin{equation}
	C_{2;23}^{[2]}(1_q^+,2^+,3^+,4_{\bar{q}}^-) = -\frac{2i\AB{4}{2-3}{1}}{\A 23^2 s_{23}}
\end{equation}
The rational contributions to the left moving primitive amplitudes for the process $q\bar{q}gg$ are
therefore given by:
\begin{align}
	R_4^{[L]}(1_q,2,3,4_{\bar{q}}) = -\frac{1}{2}\sum_{k=1}^4 C_{3;k(k+1)}^{[2]}(1_q,2,3,4_{\bar{q}})
	- \frac{s_{23}}{6} C_{2;23}^{[2]}(1_q,2,3,4_{\bar{q}})
\end{align}
Summing all of these components together can be quickly shown to match the results for four point
amplitudes \cite{Bern:qqggg,Kunszt:1l4pt}:
\begin{align}
	R_4^{[L]}(1_q^+,2^+,3^+,4_{\bar{q}}^-) &= -\frac{i\A 41\B 13}{2\A 12\A 23}
	\left( 1+\frac{2s_{12}}{3s_{23}} \right),\\
	R_4^{[L]}(1_q^+,2^+,3^-,4_{\bar{q}}^-) &= \frac{1}{2} A_4^{(0)}(1_q^+,2^+,3^-,4_{\bar{q}}^-),\\
	R_4^{[L]}(1_q^+,2^-,3^+,4_{\bar{q}}^-) &= -\frac{1}{2}\left( 1 +\frac{s_{23}}{s_{13}}
	\right)A_4^{(0)}(1_q^+,2^-,3^+,4_{\bar{q}}^-).
\end{align}
The tree amplitudes above are the well known MHV amplitudes:
\begin{align}
	A_4^{(0)}(1_q^+,2^+,3^-,4_{\bar{q}}^-) &= \frac{i\A 34^3\A 31}{\A 12\A 23\A 34\A 41},\\
	A_4^{(0)}(1_q^+,2^-,3^+,4_{\bar{q}}^-) &= \frac{i\A 24^3\A 21}{\A 12\A 23\A 34\A 41}.
\end{align}

\section{Conclusions}

We have presented a general method for extracting rational contributions to gauge-theory scattering
amplitudes using $D$-dimensional unitarity techniques \cite{Ossola:rational,Giele:dduni}. Exchanging
the $D$-dimensional cuts for four dimensional massive cuts, we then used simple complex analysis
to express the $D$-dimensional integral coefficients as contour integrals over a complex mass,
$\mu$, evaluated on a circle at infinity. Here we have gone beyond previous approaches
\cite{Britto:ddmassless,Ossola:rational,Giele:dduni} by making maximum use of the complex
behaviour of the corresponding contour integrals. We have shown that it is not necessary to
compute the
pentagon contributions explicitly, thereby reducing the computation of rational terms to the evaluation of
generalised massive cuts of box, triangle and bubble functions in the large mass-limit.  The formalism leads directly to compact analytic
expressions for the rational terms.

To demonstrate the method we have computed analytic expressions for the rational parts of one-loop
gluon amplitudes using a massive scalar loop. This has been tested by re-computing gluon helicity
amplitudes with up to six external legs. We
have also shown that the method can also be implemented numerically using a discrete Fourier transform.
The implementation was sufficient to evaluate all helicity configurations with up to six external
gluons but we leave a detailed analysis of accuracy and speed to future work.

We have also presented a simple example of how the method can be used to compute amplitudes with
massless external fermions, considering the leading colour primitive amplitude for the process
$qq\to gg$. Here we show that the rational terms can be extracted using tree amplitudes where a
massive fermion couples to a massive scalar inside the loop.

Since the procedure is based on looking at four-dimensional cuts it is reasonable to expect that one
can obtain a similar level of computational speed in comparison to calculations of the
cut-constructible terms \cite{Ellis:numuni,Berger:blackhat}, though 
further analysis of the subtraction terms is expected to be necessary. It
does have the benefit over on-shell recursive techniques that it avoids problems of unknown
factorisation properties in the complex plane and therefore applies generally to any process, in
particular to those with internal masses.

Although the present paper concentrates on applications to amplitudes with massless external
particles the main part of the procedure applies equally well to the case of massive ones. In this case
the bases for the loop momentum in each of the cuts would have to be modified slightly as described in
reference \cite{Kilgore:massuni}.

\acknowledgments

I would like to thank David Kosower and Darren Forde for valuable discussions throughout the course
of this work. I am especially indebted to Darren Forde for discussions of the fermion
amplitudes. I am also grateful to Zvi Bern for helpful
comments on this manuscript. This work
was supported by Agence Nationale de Recherche grant ANR-05-BLAN-0073-01. I would also like to thank to
Massimiliano Vincon for pointing out a number typos in a previous version of this paper.

\appendix

\section{$\mu^2$ Dependence of the Integral Coefficients \label{app:mucount}}

In this appendix we use a simple power counting argument to demonstrate the maximum power of $\mu^2$
that can appear in the $n$-point gluon amplitudes fits the format described in section
\ref{sec:dints}. Although the final result is assured for general amplitudes in any gauge theory the
following analysis sheds light on the analytic expansions of the $\inf$ expansions, showing that
only a few terms in the expansion contribute to the final rational terms. This information can then be
used for efficient computations of the boundary behaviour.
For the form to hold we must show that the leading dependence on $\mu$ appearing
in any massive box cut is $\mu^4$ and $\mu^2$ in any triangle or bubble. This property is
shown to hold in both the OPP/GKM formalism \cite{Ossola:rational,Giele:dduni} using a simple
argument from the dimension of possible tensor structures.
It is this argument that allows the parametrisation of the cut in the GKM approach and so we see
that it is related to the decompositions of the $\inf_\mu$ components of the cut given in equations
\eqref{eq:ddbox1},\eqref{eq:ddtri1} and the analogous relation for the bubble coefficients. 
In the analytic approach of Britto et al. the $\mu$-dependence can be determined through explicit
analysis of the spinor integrals \cite{Britto:ddmassless}. Here we use a slightly different
approach to justify these statements, appealing to properties of the tree level amplitudes. 

We begin by considering a general quadruple cut given by the product of four tree amplitudes.
Schematically the loop momentum can be written as,
\begin{equation}
	l(\mu^2) = \alpha_1 + \alpha_2 c + \frac{\alpha_3}{c} - \frac{\mu^2 \alpha_4}{c}
	\label{eq:boxmomC}
\end{equation}
the solution for $c$ is given by equation \eqref{eq:csol} from which it is clear that $c$ scales as
$\sqrt{\mu^2}$. It is then clear that the loop momentum given above scales as $\sqrt{\mu^2}$ in the
large $\mu$ limit.

\begin{figure}[t]
	\begin{center}
		\psfrag{1}{$p_1$}
		\psfrag{2}{$p_2$}
		\psfrag{3}{$p_3$}
		\psfrag{n}{$p_n$}
		\psfrag{n-1}{$p_{n-1}$}
		\psfrag{n-2}{$p_{n-2}$}
		\psfrag{n-3}{$p_{n-3}$}
		\psfrag{l}{$l$}
		\psfrag{l+k1n}{$l+K_{1,n}$}
		\includegraphics[width=10cm]{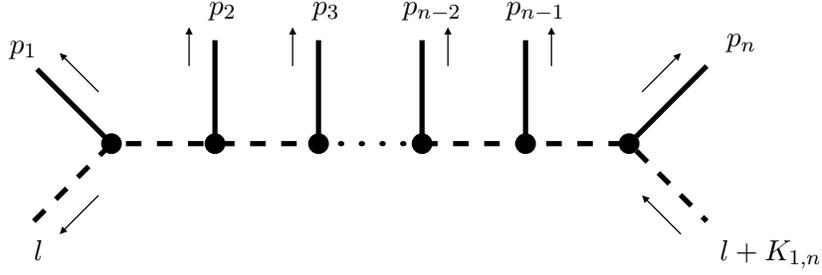}
	\end{center}
	\caption{The momentum flow of a massive scalar through a generic tree level Feynman graph.}
	\label{fig:feyncount}
\end{figure}

The next step is to consider the dependence of a generic tree amplitude on the loop
momentum. For clarity we restrict ourselves to considering a single massive scalar loop although
it should also be possible to apply a similar argument to more general cases. The tree level
amplitude can be characterised by considering a massive scalar flowing through a general Feynman
graph as shown in figure \ref{fig:feyncount}. The gluon-scalar-scalar vertex is simply proportional
to $l\cdot \varepsilon_k$ and each of the propagators will scale as,
\begin{equation*}
	\frac{1}{(l+K)^2-\mu^2} = \frac{1}{2(l\cdot K)+K^2}.
\end{equation*}
The maximum dependence on $l$ will come from graphs with the maximum number of vertices. Such graphs
will have $n$ vertices but will also have $n-1$ propagators, from which we can deduce:
\begin{equation}
	A^{(0)}_s(l) \overset{l\to\lambda l}{\overset{\lambda\to\infty}{\to}} \lambda A^{(0)}_s(l).
\end{equation}
Taken together with the scaling of the loop momentum, this is sufficient to show that the product of
four tree amplitudes scales at most as $\mu^4$ in the limit $\mu\to\infty$.

For the triangle coefficients we have to consider the expansion in $t$ in order to observe the
expected behaviour. Here the loop momentum is given as
\begin{equation}
	l(\mu^2,t) = \alpha_1 + \alpha_2 t + \frac{\alpha_3}{t} + \frac{\mu^2 \alpha_4}{t}.
	\label{eq:trimomC}
\end{equation}
We have that in any triple cut the maximum power of $t$ that can appear will be three, which follows
from the argument given above for the scaling of the tree level amplitudes. More
specifically we have maximum difference of three powers of the loop momenta between the numerator
and denominator and so we can write the boundary behaviour of a general triple cut as:
\begin{align}
	\inf_{\mu^2}[\inf_t[C_3]] &=
	\frac{P_n(a)}{P_{n-3}(b)}\\
	P_n(x) &= \sum_{m=0}^{n}\sum_{k=-n}^{n-2m} t^k\mu^{2m} x_{n-k,m}
\end{align}
In order to encode all the mass dependence in the coefficients $a,b$ we have first re-written all
explicit mass dependence in the tree amplitudes as $\mu^2=l_1\cdot l_1$. After expansion, the $t^0$
component of $C_3$ is,
\begin{align}
	&\inf_t[C_3]|_{t^0}
=\mu^2\left(
-\frac{a_{1,0}b_{2,1}}{b_{0,0}^2}
-\frac{a_{0,0}b_{3,1}}{b_{0,0}^2}
+2\frac{a_{0,0}b_{1,0}b_{2,1}}{b_{0,0}^3}
-\frac{b_{1,0}a_{2,1}}{b_{0,0}^2}
+\frac{a_{3,1}}{b_{0,0}}
\right)
\nonumber\\
&-\frac{a_{1,0}b_{2,0}}{b_{0,0}^2}
+2\frac{a_{0,0}b_{1,0}b_{2,0}}{b_{0,0}^3}
-\frac{a_{0,0}b_{3,0}}{b_{0,0}^2}
+\frac{a_{3,0}}{b_{0,0}}
-\frac{b_{1,0}a_{2,0}}{b_{0,0}^2}
+\frac{b_{1,0}^2a_{1,0}}{b_{0,0}^3}
-\frac{b_{1,0}^3a_{0,0}}{b_{0,0}^4}
\end{align}
which shows that the triangle coefficients have the expected behaviour.

Finally we use a similar argument to prove that the bubble coefficients also scale at most as $\mu^2$.
The loop momentum in this case is
\begin{equation}
	l(\mu^2,t,y) = \alpha_1 \frac{y^2}{t} + \alpha_2 \frac{y}{t} + \alpha_3 y + \alpha_4 t + \alpha_5 \frac{\mu^2}{t} + \alpha_6
	\label{eq:bubmomC}
\end{equation}
We then write down the boundary behaviour of the most general double cut integrand as a ratio of two polynomials,
\begin{align}
	\inf_\mu[\inf_t[\inf_y[[C_2]]] &= \frac{P'_n(a)}{P'_{n-2}(b)} \\
	P'_n(x) &= \sum_{m=0}^{n} \sum_{k=-n}^{n-2m} \sum_{l={\rm max}(0,-k-m)}^{n-k-2m}
	\mu^{2m} t^k y^l x_{2n-l,k+n,m}
\end{align}
Again, using the argument on the scaling of the tree-level amplitudes,
we see that the polynomial in the numerator has two powers more than the one in the numerator. The
series expansion of $C_2$ then gives us the following expression for the pure bubble cut:
\begin{align}
&\inf_t[\inf_y[C_2]]|_{t^0,y\to Y_i}=Y_2\bigg({\frac{{a_{2,2,0}}}{{b_{0,0,0}}}}
-{\frac{{a_{0,0,0}}\,{b_{2,2,0}}}{{{b_{0,0,0}}}^{2}}}
-{\frac{{a_{1,1,0}}\,{b_{1,1,0}}}{{{b_{0,0,0}}}^{2}}}
+{\frac{{a_{0,0,0}}\,{{b_{1,1,0}}}^{2}}{{{b_{0,0,0}}}^{3}}}\bigg)\nonumber\\&
+Y_1\bigg({\frac{{a_{2,2,0}}}{{b_{0,0,0}}}}
-{\frac{{a_{0,0,0}}\,{b_{2,2,0}}}{{{b_{0,0,0}}}^{2}}}
-{\frac{{a_{1,1,0}}\,{b_{1,1,0}}}{{{b_{0,0,0}}}^{2}}}
+{\frac{{a_{0,0,0}}\,{{b_{1,1,0}}}^{2}}{{{b_{0,0,0}}}^{3}}}\bigg)\nonumber\\&
+\mu^2\bigg(2\,{\frac{{a_{0,0,0}}\,{b_{2,0,1}}\,{b_{2,2,0}}}{{{b_{0,0,0}}}^{3}}}
+2\,{\frac{{a_{0,0,0}}\,{b_{1,1,0}}\,{b_{3,1,1}}}{{{b_{0,0,0}}}^{3}}}
-{\frac{{a_{2,0,1}}\,{b_{2,2,0}}}{{{b_{0,0,0}}}^{2}}}
+{\frac{{a_{4,2,1}}}{{b_{0,0,0}}}}
-{\frac{{a_{0,0,0}}\,{b_{4,2,1}}}{{{b_{0,0,0}}}^{2}}}
-{\frac{{a_{1,1,0}}\,{b_{3,1,1}}}{{{b_{0,0,0}}}^{2}}}\nonumber\\&
-{\frac{{a_{3,1,1}}\,{b_{1,1,0}}}{{{b_{0,0,0}}}^{2}}}
-{\frac{{a_{2,2,0}}\,{b_{2,0,1}}}{{{b_{0,0,0}}}^{2}}}
+{\frac{{a_{2,0,1}}\,{{b_{1,1,0}}}^{2}}{{{b_{0,0,0}}}^{3}}}
-3\,{\frac{{a_{0,0,0}}\,{{b_{1,1,0}}}^{2}{b_{2,0,1}}}{{{b_{0,0,0}}}^{4}}}
+2\,{\frac{{a_{1,1,0}}\,{b_{1,1,0}}\,{b_{2,0,1}}}{{{b_{0,0,0}}}^{3}}}\bigg)\nonumber\\&
-{\frac{{a_{0,0,0}}\,{b_{4,2,0}}}{{{b_{0,0,0}}}^{2}}}
-{\frac{{a_{1,1,0}}\,{b_{3,1,0}}}{{{b_{0,0,0}}}^{2}}}
-{\frac{{a_{1,0,0}}\,{b_{3,2,0}}}{{{b_{0,0,0}}}^{2}}}
-{\frac{{a_{2,0,0}}\,{b_{2,2,0}}}{{{b_{0,0,0}}}^{2}}}
-{\frac{{a_{2,1,0}}\,{b_{2,1,0}}}{{{b_{0,0,0}}}^{2}}}
-{\frac{{a_{2,2,0}}\,{b_{2,0,0}}}{{{b_{0,0,0}}}^{2}}}\nonumber\\&
-{\frac{{a_{3,1,0}}\,{b_{1,1,0}}}{{{b_{0,0,0}}}^{2}}}
+{\frac{{a_{2,0,0}}\,{{b_{1,1,0}}}^{2}}{{{b_{0,0,0}}}^{3}}}
+{\frac{{a_{2,2,0}}\,{{b_{1,0,0}}}^{2}}{{{b_{0,0,0}}}^{3}}}
-{\frac{{a_{3,2,0}}\,{b_{1,0,0}}}{{{b_{0,0,0}}}^{2}}}
+{\frac{{a_{4,2,0}}}{{b_{0,0,0}}}}
+6\,{\frac{{a_{0,0,0}}\,{{b_{1,0,0}}}^{2}{{b_{1,1,0}}}^{2}}{{{b_{0,0,0}}}^{5}}}\nonumber\\&
+2\,{\frac{{a_{0,0,0}}\,{b_{1,0,0}}\,{b_{3,2,0}}}{{{b_{0,0,0}}}^{3}}}
+2\,{\frac{{a_{0,0,0}}\,{b_{1,1,0}}\,{b_{3,1,0}}}{{{b_{0,0,0}}}^{3}}}
+2\,{\frac{{a_{1,1,0}}\,{b_{1,0,0}}\,{b_{2,1,0}}}{{{b_{0,0,0}}}^{3}}}
+2\,{\frac{{a_{1,0,0}}\,{b_{1,0,0}}\,{b_{2,2,0}}}{{{b_{0,0,0}}}^{3}}}\nonumber\\&
+2\,{\frac{{a_{1,0,0}}\,{b_{1,1,0}}\,{b_{2,1,0}}}{{{b_{0,0,0}}}^{3}}}
+2\,{\frac{{a_{0,0,0}}\,{b_{2,2,0}}\,{b_{2,0,0}}}{{{b_{0,0,0}}}^{3}}}
+2\,{\frac{{a_{1,1,0}}\,{b_{1,1,0}}\,{b_{2,0,0}}}{{{b_{0,0,0}}}^{3}}}
-3\,{\frac{{a_{0,0,0}}\,{{b_{1,0,0}}}^{2}{b_{2,2,0}}}{{{b_{0,0,0}}}^{4}}}\nonumber\\&
-3\,{\frac{{a_{0,0,0}}\,{{b_{1,1,0}}}^{2}{b_{2,0,0}}}{{{b_{0,0,0}}}^{4}}}
-6\,{\frac{{a_{0,0,0}}\,{b_{1,0,0}}\,{b_{1,1,0}}\,{b_{2,1,0}}}{{{b_{0,0,0}}}^{4}}}
-3\,{\frac{{a_{1,1,0}}\,{{b_{1,0,0}}}^{2}{b_{1,1,0}}}{{{b_{0,0,0}}}^{4}}}
+2\,{\frac{{a_{2,1,0}}\,{b_{1,1,0}}\,{b_{1,0,0}}}{{{b_{0,0,0}}}^{3}}}\nonumber\\&
+{\frac{{a_{0,0,0}}\,{{b_{2,1,0}}}^{2}}{{{b_{0,0,0}}}^{3}}}
-3\,{\frac{{a_{1,0,0}}\,{{b_{1,1,0}}}^{2}{b_{1,0,0}}}{{{b_{0,0,0}}}^{4}}}
\end{align}
Once we substitute the functions $Y_i$ from eq. \eqref{eq:Ys} we see that the coefficient scales as
$\mu^2$. For the triangle subtraction terms the parametrised loop momentum is the same as in the scalar
triangle coefficient but this time we are interested in the coefficients of $t^3,t^2$ and $t$:
\begin{align}
	&\inf_t[C_2^{[{\rm tri}]}]|_{t^i=T_i} = 
	 T_3\left(  
	 \frac{a_{0,0}}{b_{0,0}}
	 \right)
	+T_2\left(
	\frac{a_{1,0}}{b_{0,0}}-\frac{a_{0,0}b_{1,0}}{b_{0,0}^2}
	\right)\nonumber\\
	&+T_1\left(
	\mu^2\left(-\frac{a_{0,0}b_{2,1}}{b_{0,0}^2}+\frac{a_{2,1}}{b_{0,0}}\right)
	+\frac{a_{2,0}}{b_{0,0}}-\frac{a_{0,0}b_{2,0}}{b_{0,0}^2}-\frac{b_{1,0}a_{1,0}}{b_{0,0}^2}+\frac{b_{1,0}^2a_{0,0}}{b_{0,0}^3}	
	\right)
\end{align}
which after we use equations (\ref{eq:Ts1}-\ref{eq:Ts3}) shows that the full bubble coefficient scales as expected.

\bibliographystyle{JHEP-2}
\bibliography{ddimcuts}

\end{document}